\documentclass[12pt,USenglish]{article}
\pdfoutput=1
\usepackage{setspace}
\usepackage{cite}
\usepackage{amsmath,amssymb,amsthm,mathrsfs,bbm,subcaption}
\usepackage[utf8]{inputenc}
\usepackage{geometry}
\usepackage[USenglish]{babel}
\usepackage{verbatim}
\usepackage{prettyref}
\usepackage[normalem]{ulem}
\usepackage{graphicx}
\usepackage{setspace}
\usepackage{esint}
\usepackage{color}
\definecolor{darkgreen}{rgb}{0,0.5,0}
\definecolor{darkblue}{rgb}{0,0,0.6}
\usepackage[colorlinks=true,citecolor=darkgreen,linkcolor=darkblue,urlcolor=blue]{hyperref}

\DeclareMathOperator{\csch}{csch}

\newcommand{\bv}{{\bf v}}
\newcommand{\bk}{{\bf k}}
\newcommand{\bx}{{\bf x}}
\newcommand{\paragraphnewline}{$~$\newline}

\def\be{\begin{equation}}
\def\ee{\end{equation}}
\def\bea{\begin{eqnarray}}
\def\eea{\end{eqnarray}}
\def\ba{\begin{align}}
\def\ea{\end{align}}
\def\cO{{\cal O}}

\numberwithin{equation}{section}
\numberwithin{figure}{section}
\numberwithin{table}{section}

\begin{document}
\begin{spacing}{1.3}
\begin{flushright}

\end{flushright}

~
\vskip5mm

\begin{center} 
 {\Large \bf Overcoming obstacles in nonequilibrium holography}

\vskip10mm

Igor Novak, Julian Sonner \& Benjamin Withers\\
\vskip1em
Department of Theoretical Physics, University of Geneva, 24 quai Ernest-Ansermet, 1214 Gen\`eve 4, Switzerland
\vskip5mm

\tt{  \{igor.novak,julian.sonner, benjamin.withers\}@unige.ch}

\end{center}

\vskip10mm

\begin{abstract}
We study universal spatial features of certain non-equilibrium steady states corresponding to flows of strongly correlated fluids over obstacles. This allows us to predict universal spatial features of far-from-equilibrium systems, which in certain interesting cases depend cleanly on the hydrodynamic transport coefficients of the underlying theory, such as $\eta/s$, the shear viscosity to entropy density ratio. In this work we give a purely field-theoretical definition of the spatial collective modes identified earlier and proceed to demonstrate their usefulness in a set of examples, drawing on hydrodynamic theory as well as holographic duality. We extend our earlier treatment by adding a finite chemical potential, which introduces a qualitatively new feature, namely damped oscillatory behavior in space. We find interesting transitions between oscillatory and damped regimes and we consider critical exponents associated with these. We explain in detail the numerical method and add a host of new examples, including fully analytical ones. Such a treatment is possible in the large-dimension limit of the bulk theory, as well as in three dimensions, where we also exhibit a fully analytic non-linear example that beautifully illustrates the original proposal of spatial universality. This allows us to explicitly demonstrate how an infinite tower of discrete modes condenses into a branch cut in the zero-temperature limit, converting exponential decay into a power law tail.
\end{abstract}

\pagebreak

\pagestyle{plain}

\setcounter{tocdepth}{2}
{}
\vfill
\tableofcontents
\newpage
\section{Introduction}
Holographic duality duality as a tool for applications to strongly coupled field theories is most effective whenever one is able to identify universal quantities or mechanisms which do not depend on the precise details of the bulk theory employed, as encoded for example in the couplings appearing in the bulk action, but instead relies on universal gravitational physics, such as that associated with black hole horizons. A striking example of this universality, namely the universal ringdown of deformed horizons, has taught us a great deal about the thermalization of strongly coupled field theories with holographic duals. Linear infalling perturbations of black hole horizons in the bulk are characterized by a set of complex-frequency modes, the so-called quasinormal modes (QNM) \cite{Vishveshwara:1970zz,Press:1971wr,Horowitz:1999jd,Kovtun:2005ev}. Physically admissible modes must satisfy regularity at the future horizon and decay (usually exponentially) as a function of time. These modes are ubiquitous in our exploration of strongly coupled quantum matter via holographic duality, not least because they manifest themselves as non-analytic features in field-theory correlation functions, most commonly as poles in retarded correlation functions. By studying the dispersion relations of these poles, that is by calculating their complex frequencies $\omega(k)\in\mathbb{C}$ as a function of real momentum, much can be deduced about the relaxation dynamics of the dual field theories, including all the information about the hydrodynamic effective description of the system, that is transport coefficients, dispersion relations, and so forth.

In this paper, continuing recent work of \cite{Sonner:2017jcf}, we study a similarly universal set of modes, which govern the behavior of non-equilibrium steady states of strongly coupled field theories with holographic duals. These modes reverse the relationship of frequency and momentum described above. In other words we shall be interested in the analytic properties of correlation functions in the complex momentum plane, $k(\omega) \in \mathbb{C}$, as a function of real frequency. As will be described in detail below, these modes, which we term stationary collective modes (SCM), are generically independent and distinct from QNM. However, as we shall see, for relativistic field theories, they can be related to QNM via a procedure involving Lorentz transformations and analytic continuation.

\begin{figure}[h]
	\centering
	\includegraphics[width=0.8\columnwidth]{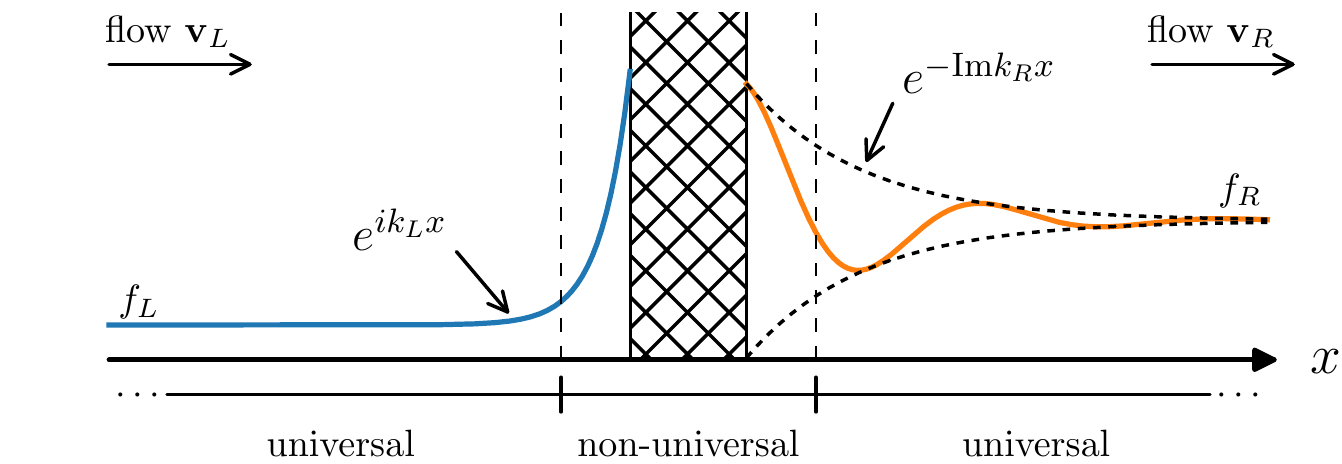}
	\caption{Schematic illustration of the nonequilibrium steady state corresponding to flow across an obstacle. Far to the left ($x\to -\infty$) and far to the right ($x\to+\infty$) the flow returns to a steady, homogeneous flow, different on each side. In the vicinity of the obstacle (hatched region) the flow is nonlinearly deformed. Connecting these two regions are a set of spatial collective modes which describe the exponential (and sometimes oscillatory) spatial relaxation back to equilibrium.}
	\label{fig:schema}
\end{figure}

The physical significance of these modes is broad and universal\footnote{Nontrivial predictions for nonequilibrium behavior based on equilibrium modes also are at the heart of the Kibble-Zurek mechanism \cite{Kibble:1976sj,Zurek:1985qw}. This analogy was pointed out previously in  \cite{Sonner:2017jcf}.}. In this work, as in  \cite{Sonner:2017jcf}, our main interest is in spatial features of non-equilibrium quantum matter, but like QNM, these modes crop up in many places, and in fact variants have already been encountered in \cite{Csaki:1998qr,Amado:2007pv,Maeda:2009wv,Khlebnikov:2010yt,Khlebnikov:2011ka,Sonner:2014tca,Blake:2014lva}. Suppose a strongly coupled field theory with a holographic dual is set up to flow across an obstacle, as might be achieved, for example, by applying a thermal gradient or an electric field. Suppose furthermore that this flow is disrupted by some obstacle, in other words that translation invariance along the flow is broken. In this case the system will arrange itself in such a way that there exists an asymptotic flow velocity, $\textbf{v}_L$, far from the obstacle and `to the left' and a generally different asymptotic flow velocity,  $\textbf{v}_R$, far from the obstacle and `to the right'. See \autoref{fig:schema} for an illustration.
In between these asymptotic regions the flow will be complicated and strongly non-linear. Nevertheless, as we show, the spatial approach towards the asymptotic regions can be universally characterized using non-analytic features of correlation functions in the complex momentum plane. 
Holographically these correspond to linear modes of the perturbed black hole which are both regular at the future horizon and which decay appropriately as one of the asymptotic spatial regions is approached. These modes are what we call spatial collective modes (SCM), as they correspond to collective excitations governing the spacelike relaxation of the strongly coupled theory.

\begin{figure}[t]
	\centering
	\includegraphics[width=0.8\columnwidth]{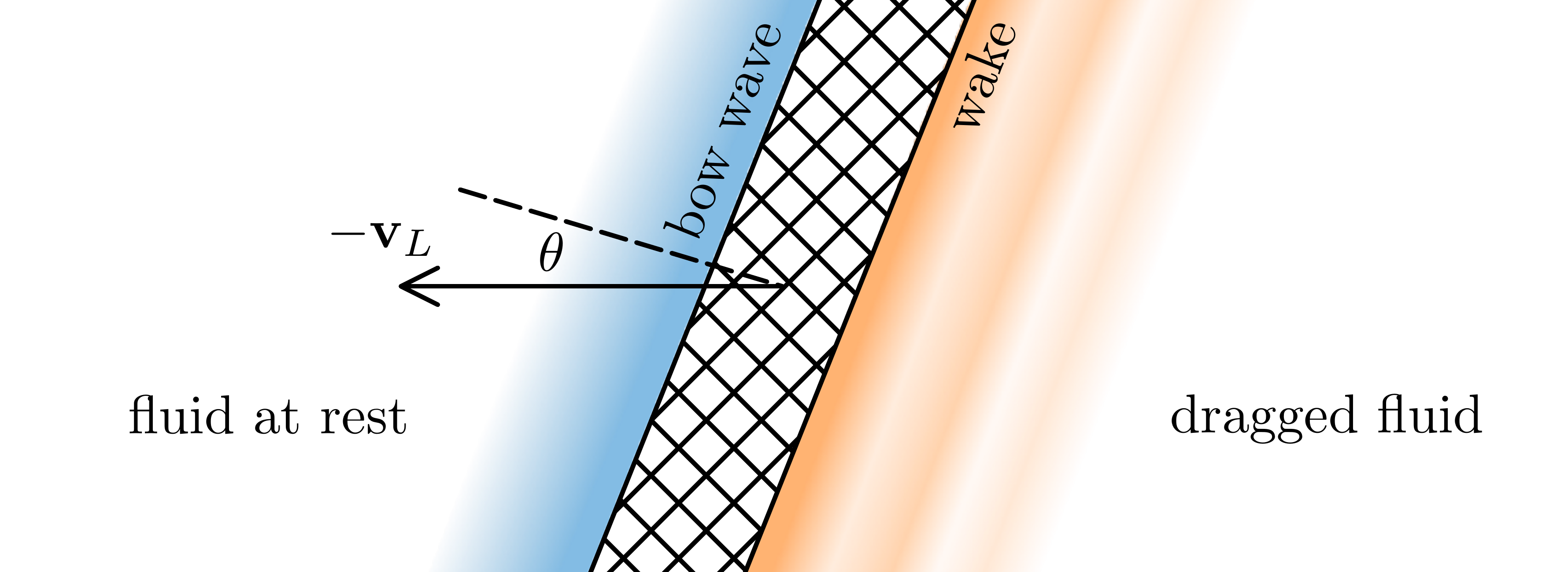}
	\caption{Schematic illustration of the nonequilibrium steady state of the type depicted in \autoref{fig:schema} in a boosted frame where the upstream fluid is at rest (here we have allowed for a fluid flow incident angle $\theta$). This corresponds to the time-dependent process of dragging a co-dimension one obstacle (hatched region) through a fluid at rest. As in \autoref{fig:schema} the fluid returns to a steady homogeneous flow far from the obstacle, and the spatial collective modes describe this process, indicated here by the spatial profiles of a bow wave and a wake.}
	\label{fig:schema_drag}
\end{figure}

In situations where the underlying theory enjoys a boost invariance, an alternative point of view on these types of steady states is provided by transforming into the frame where the fluid on the upstream side of the obstacle is at rest. In this frame, then, the physical picture is one of dragging a co-dimension one obstacle through a fluid at rest, building up a bow wave in front and leaving a wake behind, whose spatial and temporal profiles are precisely what is captured universally by the SCM described in this paper. A fixed position in the fluid at rest experiences modes which grow exponentially with time until the obstacle arrives (the bow wave), and then decay exponentially in time after the obstacle recedes (the wake). This alternative point of view is schematically depicted in \autoref{fig:schema_drag}.

We give the definition of these SCM in full generality, underlining their universal appeal, but we also find it instructive to illustrate this fact by exhibiting these modes in a number of interesting contexts, both analytically and numerically. Two such contexts in which we have in fact analytical control over the spectrum of SCM is the three-dimensional BTZ black hole, as well as the Schwarzschild black brane in a large number of dimensions $d\rightarrow \infty$. In both cases we find that the SCM are purely decaying, in other words that their defining complex momenta are in fact purely imaginary. This is not a general feature of such modes, and we go on to demonstrate that oscillatory decaying modes which also have a non-vanishing real part of the complex momentum in fact exist. We find the requisite modes in certain regimes of the dual field theory at non-zero charge density, that is in a state that is dual to a bulk Reissner-Nordstr\"om black brane. 

A second major focus of this paper is a more detailed treatment of the numerical  construction of non-equilibrium steady states dual to four-dimensional black branes with non-Killing horizons. Pursuing holographic insights into the physics of nonequilibrium steady states has proven to be a fruitful endeavour, as evidenced for example in \cite{Karch:2010kt,Sonner:2012if,Kundu:2013eba,Nakamura:2013yqa,Chang:2013gba,Bhaseen:2013ypa,Amado:2015uza, Kundu:2015qda, Spillane:2015daa, Bakas:2015hdc,Herzog:2016hob,Erdmenger:2017gdk,Figueras:2012rb,Fischetti:2012vt}, which all underline the efficiency of the holographic approach to far-from-equilibrium physics by elegantly exposing fascinating features such as emerging effective temperatures, non-equilibrium fluctuation relations, which are highly non-trivial to derive from a microscopic approach based on field theory methods. Holography serves both to reformulate the underlying non-equilibrium problem in terms of a well-posed system of partial differential equations, suitable for numerical solution, as well as to expose mechanisms and universal features through analytical insight. In this work we strive to combine both numerical and analytical insights into a quite general picture of the kind of situation described in \cite{Sonner:2017jcf}, namely stationary flows over obstacles, first introduced in \cite{Figueras:2012rb} and termed `stationary quenches'.

These solutions were employed and briefly described in our first publication  \cite{Sonner:2017jcf}, and we wish to supply a more detailed treatment of both the properties of these intriguing solutions as well as the numerical methods employed in their construction as full non-linear solutions to the bulk Einstein equations. The method is a variant of the Einstein DeTurck method of \cite{Headrick:2009pv} that applies to situations in which the generators of the bulk horizon are not Killing fields. This is essential for the non-equilibrium steady states we wish to construct, as these have broken translation invariance in the direction of the flow as mentioned above. We will also present a three dimensional example, based on the Janus solution of \cite{Bak:2011ga}, where both the non-linear and the linear analysis can be carried out fully analytically, beautifully confirming our proposal of SCM in detail, that is we are able to exhibit exactly the dominant modes governing the spatial relaxation towards the left and right asymptotic regions. In fact, we demonstrate that the black Janus solution itself should be viewed as a backreacted version of the entire tower of SCM of the three dimensional (`BTZ') black hole, which we also construct analytically. Once this is appreciated, the spectrum of SCM can in fact be recovered as an inverse Laplace transform of the non-linear solution and is seen to coincide precisely with the aforementioned tower of SCM of the three-dimensional black hole.

The large array of examples we present in this paper underscore the ubiquity of the modes proposed and defined in  \cite{Sonner:2017jcf}, and we conclude with an outline of further contexts and situations were they have promising applications. In particular as emphasised in \cite{Sonner:2017jcf} some of the dominant modes decay universally with a length proportional to $\eta/s$, presenting new opportunities to experimentally determine this ratio for strongly interacting many body systems.

The structure of this paper is as follows. In section \ref{sec.UniversalModes} we give a definition of SCM from a purely field theoretic perspective focusing on translation symmetry breaking, followed by a simple example drawn from hydrodynamic diffusion. We then work out the complete theory of SCM in charged hydrodynamics before doing the same in bulk Einstein-Maxwell theory. The remainder of section \ref{sec.UniversalModes} serves to illustrate the general definitions with a host of examples, notably a fully analytic treatment in the large-dimension limit of the bulk, as well as the three-dimensional case. In the latter we have constructed a fully non-linear examples, based on the black Janus solution, and we dedicate section \ref{sec.BlackJanus} to a detailed study of this illuminating example. Section \ref{sec.NonlinearNumerics} is dedicated to a detailed description of the numerical method employed in constructing examples of non-linear steady states where an analytical treatment is not possible. This part can be seen as a detailed companion to the original publication  \cite{Sonner:2017jcf}. The final section, \ref{sec.SummaryDiscussion} recaps the most salient features of our analysis and gives an outlook of some interesting future directions. Certain technical details throughout are relegated to two appendices to avoid overly complicating the main thrust of the paper.

\section{Universal modes in gravity and hydrodynamics}\label{sec.UniversalModes}
Our first task will be to characterise the modes which play the central role in this paper. These modes achieve for breaking spatial translations what quasinormal modes achieve for the breaking of time-translations. 

When a system is perturbed by adding a time dependent source, one can extract universal features of the late-time decay by studying certain modes in the complex frequency plane, the quasinormal modes (QNM). This is to be contrasted with our nonequilibrium steady states, where the system is perturbed along a distinguished spatial direction. Then its universal spatial relaxation at large distances from the disturbance is given by stationary collective modes (SCM), to be defined below. 
\subsection{Definition of stationary collective modes}\label{sec.DefinitionSCM}
Let us commence with a seemingly standard discussion, namely the evaluation of expectation values of operators in the interaction (`Dyson') representation of a quantum field theory. Let us suppose we have the four-momentum vector $P^\mu$ of the undeformed theory, such that a Heisenberg picture operator is given by
\be\label{eq.HeisenbergEvolution}
\Phi(x^\mu) = e^{-i P_\mu x^\mu}  \Phi e^{i P_\mu x^\mu} \,. 
\ee
Correspondingly we have the Heisenberg equation of motion for the evolution operator
\be\label{eq.HeisenbergEoM}
\partial_\mu \Phi(x) = i \left[ P_\mu, \Phi \right]\qquad \Leftrightarrow \qquad \frac{\partial U^{\rm }(x, x_0)}{\partial x^\mu} =i P_\mu U^{\rm }(x, x_0)
\ee
Usually in quantum field theory, one takes the zero component of this equation to define the time evolution of the system in question, and interprets the spatial components as giving the momentum of the system governed by its Hamiltonian evolution. For reasons that will become clear, we continue with the covariant treatment for the time being. We then have the formal solution
\be
U^{\rm }(x, x_0) = {\rm P}\exp\left(i \int^x_{x_0}P_\mu dx'{}^\mu \right)
\ee
in terms of the path-ordered exponential function, in the sense that one should interpret the integral as being a long a parametric curve $x^\mu(s)$ with $x^\mu(s_i) = x_0^\mu$ and $x^\mu(s_f) = x^\mu$ and the operators appearing being ordered in increasing order with respect to the parameter $s$. Say we now deform the theory by adding a term
\be
U^{\rm }(x, x_0) \rightarrow  U^{\rm Heis.}(x, x_0) = {\sf P}  \exp\left(i \int^x_{x_0}(P_\mu +p_\mu)dx'{}^\mu \right)\,,
\ee
where we have emphasized that we naturally obtain the evolution operator of the deformed theory in the Heisenberg picture.
It is then customary to switch to the `interaction' representation, where states evolve according to the undeformed theory, i.e. with respect to $U^{\rm }$, while operators evolve according to
\be
\Phi(x^\mu) =  U^\dagger{}^{\rm int} (x,x_0)\Phi(x_0)  U^{\rm int}(x,x_0)  \qquad  \textrm{with} \qquad U^{\rm int}(x,x_0)   = {\sf P}\exp\left( i\int_{x_0}^x p_\mu dx'{}^\mu \right)\,,
\ee
where the $x^\mu$ dependence of $p_\mu$ itself is governed only by the undeformed evolution operator $U^{\rm }(x,x_0)$.
This construction is particularly useful if we are interested in evaluating the influence of a perturbation on expectation values of the system. Let us examine this for our operator $\Phi$ above, with respect to the deformation $p_\mu$. We have
\bea
\left\langle  \Phi(x) \right\rangle &=& \left\langle  {\sf P}\exp\left( - i\int_{x_0}^x p_\mu dx'{}^\mu \right) \Phi \, {\sf P}\exp\left(  i\int_{x_0}^x p_\mu dx^\mu \right) \right\rangle \nonumber\\
&=& \left\langle  \Phi(x) \right\rangle_0  -i \int_{x_0}^x  \left\langle  \left[ \Phi(x),p_\mu(x') \right]\right\rangle_0 dx'{}^{\mu}  + \cdots
\eea
up to first order in the deformation. The subscript `$0$' on the correlator indicates that the expectation value is to be evaluated in the undeformed state. We will now describe a familiar example, where one adds an explicitly time-dependent term to the Hamiltonian of the system, before turning to the less familiar example that is the focus of this work.

\paragraph{Broken time translations}
\paragraphnewline
The familiar case usually involves the choice $p^\mu = h(t)\delta^{\mu 0}$, which leads to the well-known result
\bea
\delta \left\langle  \Phi(t,\bx) \right\rangle  &=& i \int_{-\infty}^t  \left\langle  \left[ \Phi(t,\bx),h(t') \right]\right\rangle_0 dt'\nonumber\\
&=&   \int_{-\infty}^\infty  F(t') G_R(t-t', \bx-\bx') dt'd\bx' 
\eea
where in the second line we have specialised to an $h(t)$ given by an external source $h(t)~=~\int F(t) \Phi(t,\bx)d\bx$, and we have introduced the retarded correlation function
\be
G_R(t-t',\bx-\bx') := i \theta(t-t')  \left\langle  \left[ \Phi(t,\bx)\,,\Phi(t',\bx') \right]\right\rangle_0
\ee 
The presence of the Heaviside function, which simply came from extending the integration range to $(-\infty, \infty)$, has the important consequence that the Fourier transform $\hat G_R(\omega, \bk)$ is analytic in the upper-half complex frequency plane, but may contain poles or branch cuts in the lower half plane.

Let us conclude this section by remarking that a system with the deformed Hamiltonian $H = H_0 + h(t)$ does not conserve energy, while momentum remains a conserved quantity
\be
\left[H, E \right] \neq 0, \qquad \textrm{while}\qquad \left[H, P_i \right] = 0,
\ee
which is a direct consequence of the Heisenberg equations of motion \eqref{eq.HeisenbergEoM} applied to the deformed Hamiltonian.

\paragraph{Broken spatial translations}\paragraphnewline
We now consider situations in which translation invariance is explicitly broken along a special direction ${\bf s}$, while it remains intact along the remaining spatial directions, $\bx^\|$. Let us define some coordinates by writing $\bx = (\bx\cdot \hat {\bf s}, \bx^{\|}) = (x, \bx^{\|})$. Let us thus examine the choice of deformation $p^\mu = p(x) s^\mu$, for some spacelike vector $s^\mu$ = $(0,\hat{\bf s})$, along which we assume spatial homogeneity to be broken, while all other directions remain homogeneous\footnote{An alternative picture of this situation may be given as follows: let us, for argument's sake, consider adding a term $\int F(x')\phi(t,x',\bx^\|) dx' d\bx^\|$ to the Hamiltonian. Here $x'$ is the special direction along which translation invariance is broken. Then the exponent of the evolution operator \eqref{eq.HeisenbergEvolution} formally gets a new contribution of the form
$$
Ht - {\bf P} \cdot \bx \rightarrow Ht - {\bf P}\cdot \bx +\int^t_{-\infty} F(x')\phi(t',x',\bx^\|) dx d\bx^\| dt'\,.
$$
Since we are interested in steady states, we may safely take the $t\rightarrow \infty$ limit. We now specialize to a source $F(x') = \Theta(x-x')f(x')$. At this point, it becomes more natural to actually think of the deformation as pertaining to the momentum operator, so that
$$
{\bf P}\cdot \bx \rightarrow {\bf P}\cdot \bx  - \int^x_{-\infty} p(x') dx'
$$
with $p(x') := \int F(t,x',\bx^\|) \phi(t,x',\bx^\|) dtd\bx^\|$. This is precisely the spacelike case treated above.
}.
A particular example of such a situation is given by the type of non-equilibrium steady states mentioned in the introduction, where we consider the stationary states of an interacting quantum fluid flowing over an obstacle, the obstacle being evidently the origin of the breaking of spatial translations. This also explains the `parallel' superscript which refers to the directions which are unbroken, i.e. parallel to the obstacle. This  gives\footnote{By assumption our system is in a steady state with respect to the undeformed Hamiltonian. Therefore the entire time dependence, comes from the undeformed part of the Hamiltonian.} 
\bea
\delta \left\langle  \Phi(t,\bx) \right\rangle  &=&- i \int_{-\infty}^{\bx\cdot \hat{\bf s}}  \left\langle  \left[ \Phi(t,\bx),p(x') \right]\right\rangle_0 s_\mu dx'{^\mu} \nonumber\\
&=& -i \int_{-\infty}^x  \left\langle  \left[ \Phi(t,\bx),p(x') \right]\right\rangle_0 dx'  \nonumber\\
&=&  \int_{-\infty}^\infty  F(x') G^{\scriptscriptstyle [\searrow]}(t-t', x-x',\bx^\|-\bx^\|{}'\|) dt'd\bx' \,.
\eea
In the first line the upper limit of the integral instructs us to integrate along the direction $s^\mu$ up to the spatial position $\bx\cdot \hat{\bf s}$ of the operator we are interested in measuring. In the second line we have introduced the parametrization $s_\mu dx^\mu = dx$ and written the limits on the integral accordingly. The source $F(x)$ now depends only on the spatial direction $x$. Finally, in the last step we have written the deformation as $p(x) = \int F(\bx\cdot \hat {\bf s}) \Phi (t,\bx) dt d\bx^\|$, and introduced the `decay to the right' correlator
\be
G^{\scriptscriptstyle [\searrow]}(x-x')  = -i \theta(x-x') \left\langle \left[ \Phi(t,\bx)\,,\Phi(t',\bx') \right]\right\rangle_0
\ee
where we have suppressed the dependence with respect to $t-t'$ and the remaining spatial directions for simplicity.
At first sight this looks very unfamiliar, taking the form of a `retarded' correlation function with respect to some spatial direction. We now explain why such a definition is useful. 

\begin{figure}[h]
	\centering
	\begin{subfigure}{0.5\textwidth}
	\centering
		\includegraphics[width=0.7\linewidth]{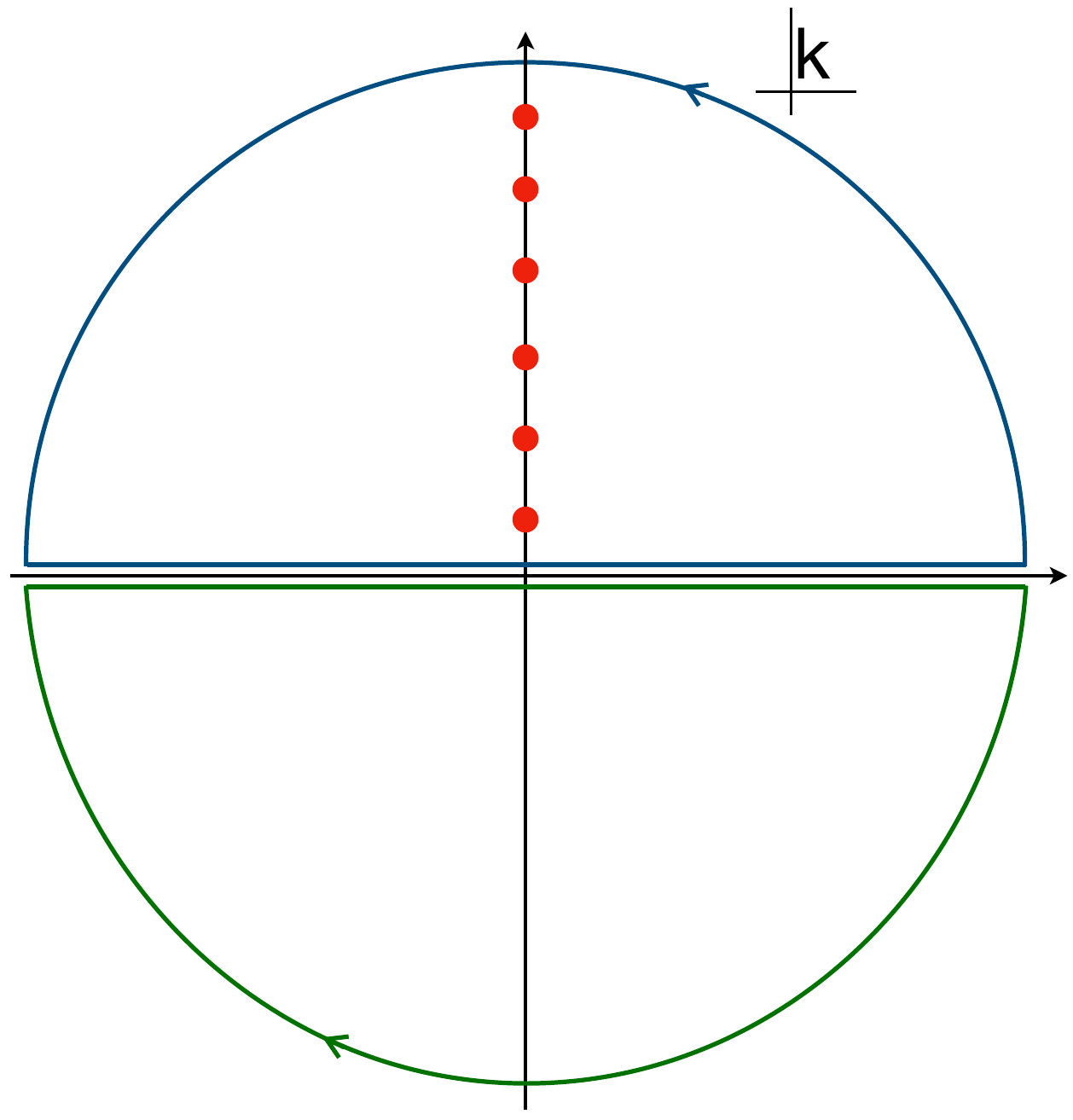}
	\end{subfigure}%
	\begin{subfigure}{0.5\textwidth}
	\centering
		\includegraphics[width=0.7\linewidth]{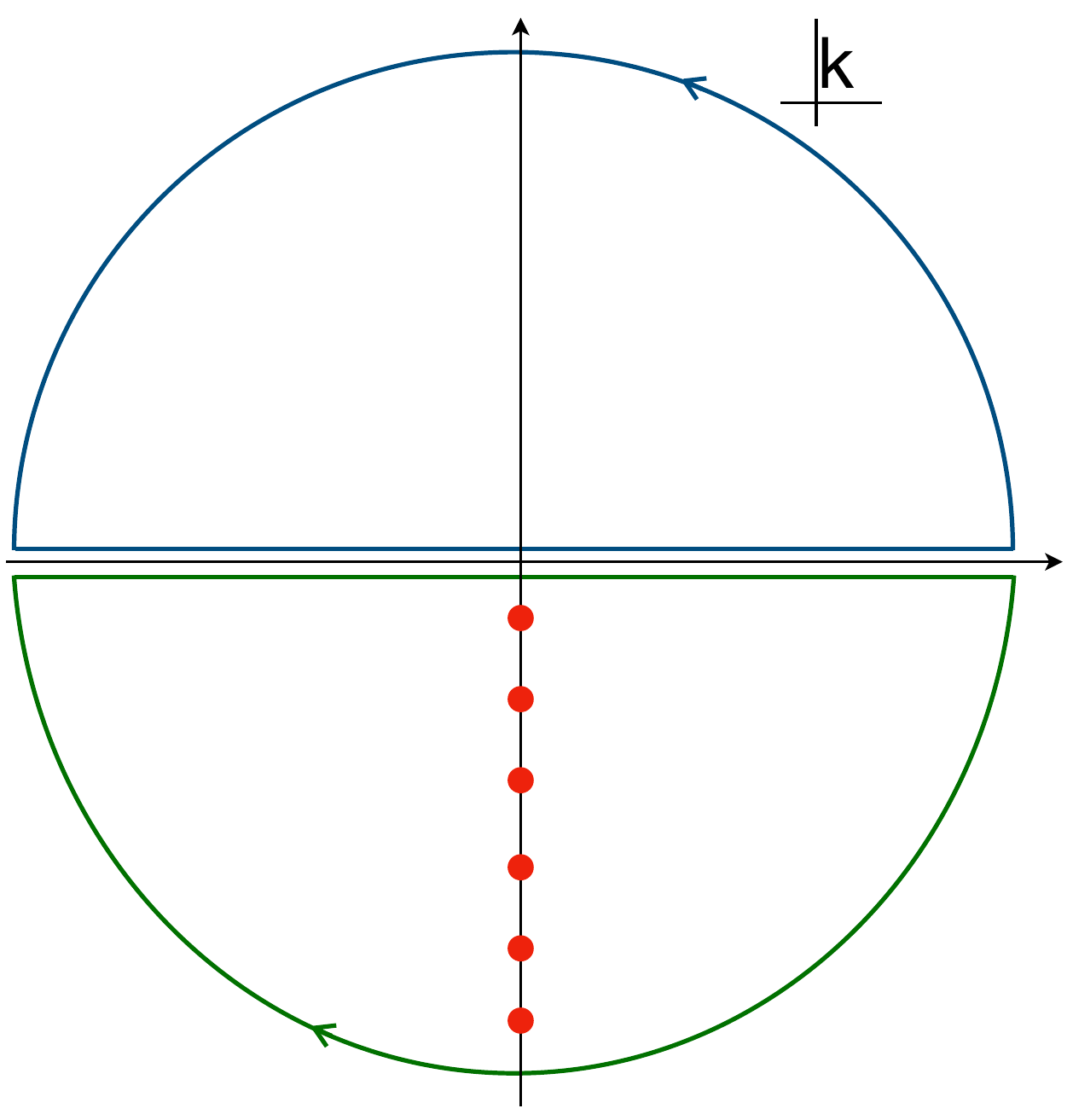}
	\end{subfigure}
	 \caption{(color figure) poles of the Fourier transform of the `decay to the right' correlation function, $\hat G^{\scriptscriptstyle [\searrow]}(k)$, and of the `decay to the left' correlation function, $\hat G^{\scriptscriptstyle [\swarrow]}(k)$, in this order, and as defined in the text. The blue contour corresponds to the region to the right of the obstacle, and the green contour to the left.\label{fig:LeftRightCorrelators}} 
\end{figure}

We first note that the Fourier transform of this object with respect to the special direction
\be
\hat G^{\scriptscriptstyle [\searrow]}(k) : = \int \frac{dx}{2\pi}  G^{\scriptscriptstyle [\searrow]}(x) e^{-ik x}
\ee
is analytic in the lower half complex $k$ plane, but may show non-analytic features, such a poles and branch cuts in the upper half complex $k$ plane. This corresponds to the situation illustrated in \autoref{fig:LeftRightCorrelators} and gives rise to modes that are decaying in the positive $x$ direction. In other words, there is an obstacle breaking spatial homogeneity in the special direction, and we enquire about the spatial profile of the expectation value of $\Phi$ to the right of the obstacle. The relevant physical solutions contributing to this quantity are therefore modes which have regular, i.e. decaying, behavior as we approach the right asymptotic region $x\rightarrow +\infty$, while they are unconstrained as $x\rightarrow -\infty$.

The analytic structure discussed here parallels what is found in the $\omega\rightarrow 0$ limit of \cite{Amado:2007pv}, and we refer the reader to this work for an insightful discussion. Illustrating further the ubiquity of SCM, these authors study modes at complex momentum in the context of equilibrium attenuation lengths of a holographic plasma, while our present work emphasizes the relevance of SCM to non-equilibrium properties of stationary states. 

The analogously defined correlator
\be
G^{\scriptscriptstyle [\swarrow]}(x-x')  = i \theta(x'-x) \left\langle \left[ \Phi(t,\bx)\,,\Phi(t',\bx') \right]\right\rangle_0
\ee
instead involves only excitations which are regular as the left asymptotic region $x\rightarrow -\infty$ is approached, while being unconstrained for $x\rightarrow +\infty$. The corresponding Fourier transform
\be
\hat G^{\scriptscriptstyle [\swarrow]}(k) : = \int \frac{dx}{2\pi}  G^{\scriptscriptstyle [\swarrow]}(x) e^{-ik x}
\ee
is analytic in the upper half complex $k$ plane, while allowing for non-trivial structure in the lower half complex plane, illustrated in \autoref{fig:LeftRightCorrelators}.  In a further analogy to the previous subsection, the system described here conserves energy, but does not conserve momentum
\be
\left[H, E \right] =0 , \qquad \textrm{while}\qquad \left[H, P_i  \right] \neq 0\,.
\ee

Without anticipating too much of our discussion in section \ref{sec.RN_SCM} below, let us briefly remark that these modes can be computed from a simple holographic prescription. There SCM are found by studying linear modes around a finite-$v$ black brane background, which are regular at the future horizon as well as regular for $x\rightarrow \infty$ for $G^{\scriptscriptstyle [\searrow]}(k)$ or regular as $x\rightarrow -\infty$ for $G^{\scriptscriptstyle [\swarrow]}(k)$, where $x$ is the distinguished direction along which translational symmetry is broken.

We will now illustrate our definitions in various different situations. We will start by a simple example, namely the hydrodynamic theory of diffusion. We will then work out the general description within a charged relativistic hydrodynamic effective theory before moving on to holography in a number of different contexts. The latter remains the focus of our paper, so we aim on giving a clear prescription of how to compute these modes in theories with holographic duals.

\subsection{A simple hydrodynamic example}
Let us furnish the definition above with a simple example. Consider a diffusive linear fluctuation, $n(t,\mathbf{x})$, obeying the equation
\be
\left(u^\mu \partial_\mu - D \Delta^{\mu\nu} \partial_\mu\partial_\nu\right)n(t,\mathbf{x}) = 0, \label{exdiff}
\ee
with diffusion constant $D\geq 0$ where $u^\mu$ is unit-normed,  timelike and future directed, which we may parameterise by a $(d-1)$-velocity, $u^\mu = \gamma (1, \mathbf{v})^\mu$, where $\gamma = (1-\mathbf{v}\cdot \mathbf{v})^{-1/2}$. The projector orthogonal to $u^\mu$ is given by $\Delta^{\mu\nu} = \eta^{\mu\nu} + u^\mu u^\nu$. With the choice $\mathbf{v} = \mathbf{0}$ \eqref{exdiff} reduces to the diffusion equation.
A defining feature of the modes we are interested in is the absence of temporal growth or decay since we are seeking the late time behaviour of a system, and this may be either steady or time-oscillatory.
A second defining feature of the collective modes is unbounded growth in one spatial direction. This is so that the mode can grow in order to match on to a source, such as an obstacle.
Let us therefore single out the coordinate $x = \bx \cdot \hat{\bf s}$ (see above), in which we permit unbounded growth, and we denote all other spatial directions by $\mathbf{x}_\parallel$ in which we do not.   For convenience we can decompose in Fourier modes in $t$ and $\mathbf{x}_\parallel$,
\be
n(t,x,\mathbf{x}_\parallel) = \int \frac{d\omega d^{d-1} k_\parallel }{(2\pi)^{d}}\hat{n}(x,\omega,\mathbf{k}_\parallel)e^{i \mathbf{k}_\parallel\cdot \mathbf{x}_\parallel - i \omega t},
\ee
and then the desired stationarity / temporal regularity condition is expressed by $\text{Im}\omega = 0$ and the desired spatial regularity condition is expressed by $\text{Im}\mathbf{k}_\parallel=0$, whilst $\hat{n}(x,\omega,\mathbf{k}_\parallel)$ may be unbounded in $x$. We make no restriction on $\mathbf{v}$ in general, but for the sake of simplicity in this example we restrict to flows which satisfy $\mathbf{v}_\parallel \cdot \mathbf{k}_\parallel = 0$. Solutions to \eqref{exdiff} are then given by,
\be
\hat{n}(x,\omega,\mathbf{k}_\parallel) = A_+ e^{\alpha_+ x} + A_- e^{\alpha_- x}
\ee
where
\be
\alpha_\pm = \frac{1}{2\gamma D}\left(v \mp \sqrt{v^2 + 4D^2 k_\parallel^2 - \frac{4 Di\omega}{\gamma}}\right) +  v i \omega.
\ee
With the given conditions on $D,k_\parallel,\omega$ one can prove that $\text{Re}\alpha_+ \leq 0$ when $v\geq 0$ and so the associated mode (with coefficient $A_+$) does not diverge as $x\to +\infty$. Thus this mode describes decay in the downstream direction. Similarly the $\alpha_-$ mode describes decay in the upstream direction. 

In the notation of section \ref{sec.DefinitionSCM}, the  $\alpha_-$ mode would appear as a pole in $\hat G^{\scriptscriptstyle [\swarrow]}(k)$, while  the $\alpha_+$ mode would appear as a pole in $\hat G^{\scriptscriptstyle ]\searrow]}(k)$.

We note that $\mathbf{k}_\parallel \neq 0$ cannot induce a nonzero $\text{Im}\alpha_-$ at $\omega = 0$, but such oscillations are seen if $\omega \neq 0$ -- a source that is oscillating in time leaves an imprinted pattern on both the upstream and downstream sides of the flow, with a wavelength set by the velocity of the flow itself.

Finally, let us remark that a similar hydrodynamic analysis for the timelike case would appear to give rise to modes that appear as poles both in the retarded and the advanced correlation function. This is actually not the case, as the diffusive nature of the hydrodynamic equation dictates a direction of time and therefore selects one or the other of the retarded or advanced correlation function depending on the sign of $D$. In other words, one cannot run the diffusion equation backwards in time without un-physically changing the sign of the diffusion constant. There is an interesting link with analyticity and causality underpinning this behavior, which we further elaborate in the discussion section.

\subsection{Charged hydrodynamics}\label{sec.HydroSCM}
Although we are mostly concerned with two and three dimensional field theories, the analysis is easily performed in $d$ space-time dimensions. As usual in hydrodynamics we begin by writing down the stress tensor and current in a derivative expansion
\begin{align}
 T_{\mu\nu}&=T^{(0)}_{\mu\nu} + \Pi^{(1)}_{\mu\nu} + O(\partial)^2, \\
 J_{\mu}&= J^{(0)}_\mu + J^{(1)}_\mu +  O(\partial)^2
\end{align}
where, up to first order, we have the Landau frame expressions
\begin{align}
 T^{(0)}_{\mu\nu} &= \varepsilon u_{\mu}u_{\nu} + p\Delta_{\mu\nu}, \\
 \Pi^{(1)}_{\mu\nu} &= -\eta\sigma_{\mu\nu}-\zeta\Delta_{\mu\nu}\partial\cdot u,\\
 J^{(0)}_\mu &= nu_{\mu},\\
 J^{(1)}_\mu &= -\sigma T\Delta_\mu{}^{\nu}\partial_{\nu}\left( \frac{\mu}{T} \right) \label{eq.current}\,.
\end{align}
Here the shear tensor is given by
\begin{equation}
 \sigma_{\mu\nu}=2\Delta_{\mu}\,^{\rho}\Delta_{\nu}\,^{\sigma}\left( \partial_{(\rho}u_{\sigma)}-\frac{1}{d-1}\eta_{\rho\sigma}\partial\cdot u \right)\,,
\end{equation}
where $\varepsilon$ is the energy density, $p$ is the pressure, $u^{\mu}$ is a timelike unit-normalised $d$-velocity field while $n$ is the charge density. The tensor $\Delta^{\mu\nu}=\eta^{\mu\nu} + u^{\mu}u^{\nu}$ projects orthogonally to $u^{\mu}$. The quantities $\eta$ and $\zeta$ are the shear and bulk viscosities, and $\sigma$ is the charge conductivity. The energy-stress tensor and the current are subject to conservation equations:
\begin{align}\label{conservation}
 \partial_{\mu}T^{\mu\nu} &= 0, \\
 \partial_{\mu}J^{\mu} &=0.
\end{align}
These conservation laws give rise to $d+1$ equations for the $d+2$ unknowns contained in $u^\mu, \varepsilon, p, n$. Consequently, this system of equations still needs to be closed by specifying an equation of state $p(T,\mu)$. This equation of state depends on the physical system under consideration, but for the majority of this paper we are interested in holographic field theories, which are conformal in the UV. The required general conformal equation of state, and the specific example of the conformal equation of state for the Reissner-Nordstr\"om solution can be found in appendix \ref{app.EqnOfState} together with its associated transport coefficients.

In order to find the collective modes, we solve the conservation equations for linear perturbations about a long-range stationary state characterised by $\varepsilon$, $p$, $n$ and a $(d-1)$ velocity $\bv$, such that $u^{\mu}=\gamma(1,\bv)$ where $\gamma =\frac{1}{\sqrt{1-\bv\cdot\bv}}$. We then consider time-independent perturbations of the form
\begin{align}
 \varepsilon\left( x^{\mu} \right) &= \varepsilon + \delta\varepsilon\, e^{ik_{\sigma}x^{\sigma}}, \\
 p\left( x^{\mu} \right) &= p + \delta p\, e^{ik_{\sigma}x^{\sigma}}, \\
 n\left( x^{\mu} \right) &= n + \delta n\, e^{ik_{\sigma}x^{\sigma}}, \\
 u^{\mu}\left( x^{\mu} \right) &= u^{\mu} + \delta u^{\mu}\, e^{ik_{\sigma}x^{\sigma}}.
\end{align}
The modes we are interested in are time independent in the laboratory frame, namely the frame in which we have a steady state, that is they have $k_{\mu}=(0,\bk)$ in the frame where $u^{\mu}=\gamma(1,\bv)$. Inserting these perturbations into the energy-stress tensor, we find
\begin{align}
 \delta T^{\mu\nu}_{(0)} &=\left( u^{\mu}u^{\nu}\delta\varepsilon + \Delta^{\mu\nu}\delta p +2(\varepsilon + p)u^{(\mu}\delta u^{\nu)} \right)e^{ik_{\sigma}x^{\sigma}}, \\
 \delta\Pi^{\mu\nu}_{(1)} &=\left( -2\eta\Delta^{\rho(\mu}\delta u^{\nu)} + \left( \frac{2\eta}{d-1} - \zeta \right)\Delta^{\mu\nu}\delta u^{\rho} \right)ik_{\rho}e^{ik_{\sigma}x^{\sigma}}.
\end{align}
The conservation equations (\ref{conservation}) then give
\begin{eqnarray}\label{delta T}
 ik_{\mu}\delta T^{\mu\nu}_{(0)} + ik_{\mu}\delta\Pi^{\mu\nu}_{(1)} + O(k^3) &=& 0,\label{eq.stressEnergyConservation}\\
 ik_\mu J^\mu_{(0)} +  ik_\mu J^\mu_{(1)}+ O(k^3) &=&0\,.
\end{eqnarray}
Finally let us use the equation of state to define the suceptibilities
\begin{equation}
 \alpha_1 = \left(\frac{\partial\mu}{\partial\varepsilon}\right)_{n} - \frac{\mu}{T}\left(\frac{\partial T}{\partial\varepsilon}\right)_{n}, \quad \alpha_2=\left(\frac{\partial\mu}{\partial n}\right)_{\varepsilon} - \frac{\mu}{T}\left(\frac{\partial T}{\partial n}\right)_{\varepsilon}
\end{equation}
and
\begin{equation}\label{eq.betaSuscept}
 \beta_1=\left(\frac{\partial p}{\partial\varepsilon}\right)_n , \quad \beta_2=\left(\frac{\partial p}{\partial n}\right)_{\varepsilon}\,.
\end{equation}
We now have all formulae and definitions in place to determine our SCM. Once an equation of state is specific these quantities can be found explicitly, see for example \ref{app.EqnOfState} for the case of the conformal equation of state in $d$ dimensions. In the following, it is most convenient to split the analysis into different channels, according to whether the velocity field perturbation is transverse or longitudinal with respect to the momentum.

\subsubsection{Transverse channel}\label{sec:hydrotrans}
We start with the transverse channel, that is velocity field perturbations $\delta u$, such that $k\cdot \delta u =0$. We find
\begin{eqnarray}
(k\cdot u) \delta n - i \sigma \Delta_{k^2} \left( \alpha_1 \delta \varepsilon + \alpha_2 \delta n\right) = 0\\
(k\cdot u) u^\nu \delta \varepsilon + k_\mu \Delta^{\mu\nu} \left(  \beta_1 \delta\varepsilon + \beta_2\delta n  \right) + \left(  \varepsilon + p\right)(k\cdot u) \delta u^\nu - i \eta \Delta_{k^2} \delta u^\nu = 0 \label{eq.umuTransverse}
\end{eqnarray}
where here and below we use the shorthand
\be
\Delta_{k^2} = k_\mu k_\nu \Delta^{\mu\nu}\,.
\ee
Let us first examine the case when $k\cdot u =0$, i.e. when the momentum of the perturbation, in addition to being transverse to $\delta u$, is also transverse to the background flow velocity. First we note that the stress-tensor conservation, contracted with $u$ gives the relation
\be
(k\cdot u)\delta \varepsilon = - (\varepsilon + p) k\cdot \delta u
\ee
i.e. that $k\cdot u = 0$ implies that the perturbation is transverse $k\cdot \delta u =0$. Assuming this is the case, the charge conservation equation gives
\be
\delta n = -\frac{\alpha_1}{\alpha_2}\delta\varepsilon
\ee
which when plugged back into the stress tensor conservation tells us that 
\be
\delta u^\mu = \frac{\alpha_2\beta_1 - \alpha_1\beta_2}{i\eta\alpha_2}\frac{k^\mu }{k^2}\delta\varepsilon\,.
\ee
However, this contradicts our assumption that $\delta u$ is normal to $k$, implying that our system of equations has no solution unless the quantity in the numerator above vanishes. As we will see below, this combination of susceptibilities is related to the charge diffusion constant, which is generically non-zero. We conclude that no non-trivial mode for $k\cdot u = 0$ exists. We may thus proceed with our analysis, assuming $k\cdot u \neq 0$.

In order to make further progress, we start by projecting Eq. (\ref{eq.umuTransverse}) onto $u^\mu$, keeping in mind that the normalization condition on the flow velocity implies that $u\cdot \delta u  =0$. We immediately find that $\delta \varepsilon = 0$. Projecting (\ref{eq.umuTransverse}) onto $k$, we discover in addition that $\delta n=0$. Notice that this follows from the fact that $\Delta_k^2 \neq 0$, since both contributing terms are positive definite for our choice of $k^\mu$ in the lab frame. We are thus left with the nontrivial condition
\be
\left[ (\varepsilon + p) (k\cdot u) - i \eta \Delta_{k^2}\right]\delta u^\mu =0
\ee
Writing
\be\label{eq.LabFrameVelocity}
u^{\mu}=\gamma(1,\bv) \,,\qquad k^\mu = (0,\bk)^\mu\,,
\ee
so that $\bv\cdot\bk = vk \cos\theta$, we conclude that there is a non-trivial transverse mode with dispersion relation
\be
 (\varepsilon + p)\frac{kv\cos\theta}{\sqrt{1-v^2}} - i\eta k^2\frac{1-(v\sin\theta)^2}{1-v^2} + O(k^3) = 0.
\ee
for which
\be
\delta\varepsilon = \delta n =0\,,\qquad \delta u^\mu \neq 0 \,,\qquad (\textrm{with}\quad k\cdot \delta u =0)\,.
\ee
This concludes our analysis of the transverse channel. Let us now turn to the longitudinal channel.
\subsubsection{Longitudinal channel}\label{sec:hydrolong}
A longitudinal perturbation satisfies $\delta u^{\mu} = \delta u_{L}\Delta^{\mu\nu}k_{\nu}/k$, where $k = \sqrt{k_\mu k^\nu}$ is the norm of the spacelike momentum. As we argued above, without loss of generality we can take $k\cdot u\neq 0$ in the longitudinal channel and we do so from now on. There are three independent modes in this channel, which we determine as follows. We first substitute the form of $\delta u_L $ into the dynamical equations (\ref{delta T}). These then take the form of a scalar equation (the current conservation equation) and a vector equation (the stress tensor conservation equation). We may use the equation of state to eliminate $\delta p$ from these, and finally project the vector equation first onto $k_\mu$ and then onto $u_\mu$. This results in the following system of three linear equations
\bea\label{eq.longitudinalSector}
\left(\begin{array}{ccc}
k\cdot u  & 0 & (\varepsilon+p) \frac{\Delta_k^2}{k}\\
-i \sigma \Delta_k^2 \alpha_1 & k\cdot u - i \sigma \Delta_k^2 \alpha_2 \ & n \frac{\Delta_k^2}{k}\\
\frac{\Delta_k^2}{k}\beta_1 & \frac{\Delta_k^2}{k}\beta_2 & \frac{\Delta_k^2(\varepsilon+p)}{k^2}\left(k\cdot u - i \Delta_k^2 \gamma_s\right)
\end{array}\right)
\left(\begin{array}{c}
\delta \varepsilon \\ 
\delta n \\ 
\delta u_L
\end{array} \right)+ 
\left(\begin{array}{c}
0 \\ 
O(k)^3 \\ 
O(k)^3
\end{array} \right)= 0\,,
\eea
expanded in small momentum $k$ and where we have defined
\be
\gamma_s\equiv\frac{\frac{d-2}{d-1}2\eta + \zeta}{\varepsilon+p}.
\ee
Note that the first of these equations corresponds to the conservation of energy and is exact in $k$. The second of these is the charge conservation equation and is correct to second order in $k$ with order $k^3$ terms truncated. The remaining equation is correct to second order in $k$. A physical mode, that is a hydrodynamic SCM, corresponds to a non-trivial solution of this linear equation, in other words an eigenmode with eigenvalue zero. Each eigenvalue is a polynomial in $k$ and the vanishing of this polynomial gives the dispersion relation of a non-trivial longitudinal SCM admitted by the system of equations \eqref{conservation}.

We shall now explicitly construct the eigenvalues and associated eigenmodes for the system \eqref{eq.longitudinalSector}, working order-by-order in $k$. Specifically, denoting the matrix multiplying $(\delta \varepsilon, \delta n,  \delta u_L)^\text{T}$ in \eqref{eq.longitudinalSector} by $M$, we wish to solve, 
\be
M \cdot V_I = \lambda_I V_I \label{MVemode}
\ee 
where $I=1,2,3$ labels the eigenmode. As in the transverse channel we have $\bv\cdot \bk = v k \cos\theta$. Next we expand as follows in powers of $k$,
\bea
M &=&  M^{(1)}  k+ M^{(2)} k^2+O(k)^3\\
V_I &=& V_I^{(0)} + V_I^{(1)} k +O(k)^2\\
\lambda_I &=&  \lambda_I^{(1)} k+ \lambda_I^{(2)} k^2+O(k)^3\\
v\to v_I &=& v_I^{(0)} + v_I^{(1)} k + O(k)^2
\eea
Then we solve the eigenmode equation \eqref{MVemode}, where at each order setting $\lambda^{(n)}_I = 0$ determines $v_I^{(n-1)}$ for the mode in question labelled by $I$. Note that in the results that follow we have fixed a freedom to shift $V_I^{(1)}$ by a multiple of $V_I^{(0)}$, and we have done so simply by looking for the most compact presentation.

At $\cO(k)$ we have the following eigenmode problem,
\be
M^{(1)}\cdot V_I^{(0)} = \lambda_I^{(1)}V_I^{(0)}
\ee
which we solve directly to obtain $V^{0}_I$ and $\lambda_I^{(1)}$.  Each solution to this leading order problem picks a different physical mode, with $\lambda_I^{(1)} = 0$ corresponding to the leading order part of its dispersion relation. In the following we shall treat each in turn, denoting $c_\theta \equiv \cos\theta$ and $s_\theta \equiv \sin\theta$, for compactness.

\paragraph{Diffusive mode}\paragraphnewline
In this case the fluctuations, up to first order in the momentum expansion, have dispersion relation and eigenmodes, given by
\bea
v &=&  i\frac{D_{\alpha\beta}}{c_\theta}  k + O(k)^2\\
\{\delta \varepsilon, \delta n,  \delta u_L\}^{(0)} &=& \left\{-\beta_2,\beta_1,0\right\}\\
\{\delta \varepsilon, \delta n,  \delta u_L\}^{(1)} &=& \left\{0,0,i\beta_2 \frac{D_{\alpha\beta}}{\varepsilon + p}\right\}
\eea
Where we have defined a diffusion constant and the speed of sound,
\be
D_{\alpha\beta} =  \sigma\frac{\alpha_2 \beta_1 - \alpha_1\beta_2}{c_s^2}\,,\qquad  c_s^2 \equiv \beta_1 + \frac{n}{\varepsilon+p}\beta_2\,.
\ee
We may think of this mode as primarily accounting for charge density perturbations since, for example, in a conformal theory ($\beta_2=0$, see appendix \ref{app.EqnOfState}) it consists solely of $\delta n$, even at nonzero background velocities. In more general theories, we still have a diffusive (purely imaginary) dispersion relation for charge fluctuations, which are, however, coupled to non-trivial energy and velocity fluctuations.

\paragraph{Sound-like modes}\paragraphnewline
For sound-like modes the fluctuations, up to first order in the momentum expansion, take the form
\bea\label{eq.ChargedSoundMode}
v &=& \pm v_0 + i \Gamma \frac{\left(1-s_\theta^2v_0^2\right)^2}{2c_\theta}\gamma_0 k + O(k)^2\\
\{\delta \varepsilon, \delta n,  \delta u_L\}^{(0)} &=&  \left\{\left(1-s_\theta^2v_0^2\right)\gamma_0^2(\varepsilon + p),\left(1-s_\theta^2v_0^2\right)\gamma_0^2 n,  \mp c_\theta\gamma_0 v_0\right\}\\
\{\delta \varepsilon, \delta n,  \delta u_L\}^{(1)} &=&\gamma_0^3 \left(1-s_\theta^2v_0^2\right)\left\{0, \pm i  c_\theta v_0 \frac{\Gamma-\gamma_s}{\beta_2(\varepsilon+p)}, - \frac{i}{2} \gamma_0\left(1-(1+c_\theta^2)v_0^2\right) \Gamma\right\}\nonumber\\
\eea
where we have introduced a second diffusion constant
\be
\Gamma\equiv\gamma_s + \frac{\sigma\beta_2\left(\alpha_1+\frac{n}{\varepsilon+p}\alpha_2\right)}{c_s^2},
\ee
and where $\gamma_0 = \frac{1}{\sqrt{1-v_0^2}}$ for a speed of sound modified by the angle of incidence,
\be
v_0 = \frac{c_s}{\sqrt{ c_s^2s_\theta^2 + c_\theta^2}}.
\ee
Note that this mode does not contain any $\delta n$ component for neutral backgrounds $\mu = n = 0$. 

Let us now further explain why we used the terminology {\it diffusive} and {\it sound like} for our SCM, which will also show that $\Gamma$ is related to conventional sound attenuation.

\paragraph{Relation to conventional hydrodynamic modes}\paragraphnewline
Our SCM in the laboratory frame are defined to have vanishing imaginary part of the frequency and appear for complex values of momentum. For simplicity we here restrict to zero frequency. More generally one can also consider SCM with nonzero frequency similar to the equilibrium modes studied in \cite{Amado:2007pv}. Due to the underlying Lorentz symmetry it is possible, at least formally, to transform these modes back into the rest frame of the fluid. There they can be analytically continued into modes which satisfy dispersion relations, more conventionally associated with hydrodynamic modes, or in the gravity case, hydrodynamic quasinormal modes. 
In this way the diffusive SCM is related to ordinary charge diffusion in the fluid rest frame where this mode obeys the dispersion relation $\omega=-iD_{\alpha\beta}q^2$. Similarly the sound-like SCM is related via boost and analytic continuation to a standard sound mode with dispersion relation $\omega = \pm c_sq - i\frac{\Gamma}{2}q^2$.
We emphasize once more that this procedure relies on the Lorentz symmetry of the underlying theory, as well as analytic continuation, and will fail for a non-relativistic theory. In general, i.e. in the absence for Lorentz symmetry, the SCM considered in this paper, are physically distinct from and independent of quasinormal modes.

 \subsection{Neutral hydrodynamics}
 Here we review briefly the special case of a neutral fluid, which we discussed in a previous publication \cite{Sonner:2017jcf}. In the present context the results of \cite{Sonner:2017jcf} can be recovered simply as the limit of zero charge density, $n\rightarrow 0$, of the charged case. For completeness we repeat the salient equations here.  \cite{Sonner:2017jcf} did not consider charge density fluctuations, as there was no bulk gauge field. In that case, only the sound-like mode arises, whose dispersion relation indeed corresponds to the $n\rightarrow 0$ limit of \eqref{eq.ChargedSoundMode}.
 \subsubsection{Transverse channel}
 The analysis for the transverse mode proceeds as for the charged case above, resulting in the un-modified transverse mode dispersion relation
 \be
 (\varepsilon + p)\frac{kv\cos\theta}{\sqrt{1-v^2}} - i\eta k^2\frac{1-(v\sin\theta)^2}{1-v^2} + O(k^3) = 0\,,\label{neutraltransverse}
 \ee
 which was given previously in \cite{Sonner:2017jcf}. 
 
\subsubsection{Longitudinal channel}

\paragraph{Diffusive mode}\paragraphnewline
We again have a diffusive mode with dispersion relation $v =  i\frac{D_{\alpha\beta}}{c_\theta}  k + O(k)^2$, and mode structure
 \bea
\{\delta \varepsilon, \delta n,  \delta u_L\}^{(0)} &=& \left\{-\beta_2,\beta_1,0\right\}\\
\{\delta \varepsilon, \delta n,  \delta u_L\}^{(1)} &=& \left\{0,0,i\beta_2 \frac{D_{\alpha\beta}}{\varepsilon + p}\right\}\,.
\eea
The constant $D_{\alpha\beta}$ is defined as before, and the speed of sound reduces to $c_s^2 \equiv \beta_1$. This mode is primarily a charge diffusion mode, and was consequently not considered in the analysis of the neutral fluid in \cite{Sonner:2017jcf}.\vskip1em

\paragraph{Sound-like modes}\paragraphnewline
Finally, we have the sound mode with dispersion relation $v = \pm v_0 + i \Gamma \frac{\left(1-s_\theta^2v_0^2\right)^2}{2c_\theta}\gamma_0 k + O(k)^2$ and mode structure
\bea\label{eq.NeutralSoundMode}
\{\delta \varepsilon, \delta n,  \delta u_L\}^{(0)} &=&  \left\{\left(1-s_\theta^2v_0^2\right)\gamma_0^2(\varepsilon + p),0,  \mp c_\theta\gamma_0 v_0\right\}\\
\{\delta \varepsilon, \delta n,  \delta u_L\}^{(1)} &=&\gamma_0^3 \left(1-s_\theta^2v_0^2\right)\left\{0,  0 , - \frac{i}{2} \gamma_0\left(1-(1+c_\theta^2)v_0^2\right) \Gamma\right\}
\eea
We can solve the dispersion relation for $k$ to obtain
\be
k = -i \frac{\varepsilon+p}{\frac{d-2}{d-1}\eta + \frac{1}{2}\zeta} \frac{\sqrt{1-v_0^2}\cos\theta}{(1-(v_0\sin\theta)^2)^2}(v\mp v_0) + O(k)^2
\ee
which is the form given in  \cite{Sonner:2017jcf}. We shall now move on to a different arena in which we can characterise our SCM in a detailed fashion, namely holography. Since this is a microscopic theory going beyond the hydrodynamic limit, we will also be able to illustrate SCM that are not captured by a hydrodynamic effective theory. 
\subsection{Reissner-Nordstr\"om AdS}\label{sec.RN_SCM}
Since we would like to compute modes pertaining to steady states of a finite-temperature finite charge system, the appropriate bulk theory is Einstein-Maxwell.
We start from such a theory in the bulk in the conventions of \cite{Hartnoll:2009sz} 
\begin{equation}
S_{\textrm{bulk}} = \int d^{d+1}x \sqrt{-g} \left[ \frac{1}{2\kappa^2}\left( R + \frac{d(d-1)}{L^2} \right) - \frac{1}{4\tilde{g}^2} F^2\right].\label{EMaction}
\end{equation}
Here $\tilde{g}$ is the Maxwell coupling and $\kappa = 8\pi G_N$. The relevant background is the Reissner-Nordstr\"{o}m black hole in AdS, with line element given by
\begin{equation}
ds^2 = \frac{L^2}{z^2}\left( -f(z)dt^2 + \frac{dz^2}{f(z)} + dx^{i}dx^{i} \right),
\end{equation}
where $L$ is the AdS radius and
\begin{equation}
f(z) = 1 - \left(1+\frac{z^{2}_{h}\mu^2}{\tilde{\gamma}^2} \right)\left( \frac{z}{z_{h}} \right)^{d} + \frac{z^{2}_{h}\mu^2}{\tilde{\gamma}^2}\left( \frac{z}{z_{h}} \right)^{2(d-1)}.\label{RNfdef}
\end{equation}
In the expression for $f(z)$ we have introduced the dimension-dependent constant $\tilde{\gamma}$, defined as
\begin{equation}
\tilde{\gamma}^{2} = \frac{(d-1)\tilde{g}^2 L^2}{(d-2)\kappa^2}.
\end{equation}
In these coordinates the conformal boundary is located at $z=0$, and we denote the position of the horizon by $z_h$.
The scalar potential is nontrivial and given by
\begin{equation}
A_{t} = \mu \left[1-\left( \frac{z}{z_{h}} \right)^{d-2}  \right],
\end{equation}
and the Hawking temperature by the expression
\begin{equation}
T = \frac{1}{4\pi z_{h}}\left( d - \frac{(d-2)z^{2}_{h}\mu^2}{\tilde{\gamma}^2} \right).
\end{equation}
Now, we boost along the planar horizon direction by $u_{\mu}$. Writing the boosted metric in ingoing Eddington-Finkelstein coordinates we have
\begin{equation}
ds^{2}_{\textrm{boosted}} = \frac{L^2}{z^2}\left( -f(z)(u_{\mu}dx^{\mu})^2 + 2u_{\mu}dx^{\mu}dz + \Delta_{\mu\nu}dx^{\mu}dx^{\nu}\right),
\end{equation}
where
\begin{equation}
\Delta_{\mu\nu} = \eta_{\mu\nu} + u_{\mu}u_{\nu} \quad \textrm{and} \quad u_{\mu} = \frac{1}{\sqrt{1-\mathbf{v}^2}}(1,\mathbf{v}).
\end{equation}
The Greek indices $\mu$, $\nu$ refer to boundary coordinates, i.e. $\mu,\nu = v, x, y$, while the Latin indices encompass all of the coordinates. To construct the spatial collective modes from the bulk perspective, we linearly perturb the metric. We take the following ansatz for the perturbations
\begin{align}\label{perturbations}
\delta g_{ab}(z, x^{\mu}) dx^{a}dx^{b} & =  h_{\mu\nu} (z) e^{ik_{\sigma}x^{\sigma}}dx^{\mu}dx^{\nu}, \\
\delta A_{a}(z, x^{\mu}) dx^a & =  H_{\mu}(z) e^{ik_{\sigma}x^{\sigma}} dx^\mu.
\end{align}
There are no perturbations that involve the radial direction $z$ by our choice of axial gauge. 

For sake of specificity in the calculations, which we eventually perform numerically, we boost the metric in the $x$-direction
\begin{equation}
\mathbf{v} = (0,v_x,0)
\end{equation}
The perturbations (\ref{perturbations}) give rise to a set of coupled ODEs in $z$ in the Einstein and Maxwell equations. It is most convenient to organise them into a transverse and longitudinal channel with respect to the obstacle\footnote{For a detailed covariant treatment of this decomposition, for a general incident angle, see the Supplementary Material of \cite{Sonner:2017jcf}}. The $h_{xy}$, $h_{vy}$ and $H_{y}$ belong to the transverse channel, whereas $h_{vv}$, $h_{xx}$, $h_{yy}$ and $H_v$, $H_x$ belong to the longitudinal channel. The two channels cannot mix and thus form two sets of mutually decoupled linear differential equations.

We must also specify boundary conditions. In analogy with a similar procedure for calculating black hole quasinormal modes, we require regularity on the future event horizon, and in addition regularity in one of the two asymptotic directions \cite{Sonner:2017jcf}. For a detailed discussion, the Reader may refer back to section \ref{sec.DefinitionSCM}.

We have not been able to solve the equations analytically\footnote{Although, see section \ref{sec.BTZSCM} for the AdS$_3$ case, where analytical solutions can in fact be obtained.}. We therefore solve these ODEs numerically, utilizing a double-sided shooting method. We specify input boundary and horizon data and then integrate the equations on one side from the boundary, and on the other from the horizon to a common mid-point. In order to have a solution the numerical values obtained by integrating from both sides must match at the mid-point for the functions and whenever a component equation is of second order, also their first derivatives. This is achieved by finding suitable values of the input data via a Newton-Raphson routine. The routine changes the values we initially specify incrementally, until the mid-point solutions are a match up to a certain precision we specify. In our case, for most solutions, the threshold for an acceptable solution was chosen such that the absolute value of the difference at the mid-point was less than $10^{-8}$.

In the concrete calculations we report on below, we choose $d=3$, that is we work in AdS$_4$ and to simplify things further, we choose $L=z_{h}=1$ and units in which $\tilde{g}^2=2\kappa^2 = 1$. 

\subsubsection{Transverse channel}
The ODEs in the transverse channel are given by the following expressions,
\begin{align}
ik_x\gamma h_{vy}+ \mu v\gamma H_{y} +\frac{(f-1)v\gamma^2}{z^2}\left( z^2 h_{vy}\right)' + \frac{1+(f-1)\gamma^2}{z^2}\left( z^2 h_{xy}\right)' = 0, \\
\mu\gamma\left( z^2(h_{vy}+vh_{xy}) \right)' + k^{2}_{x}\gamma^2 H_{y} + 2ik_{x}v\gamma H'_{y} - \left(fH'_{y} \right)' =0,\\
\frac{ik_x \mu v \gamma^2}{f}H_{y} + \mu\gamma H'_{y} + \frac{k^{2}_{x}\gamma^2}{f}h_{vy} + \frac{2ik_x v \gamma}{zf}\left( z h_{vy} \right)'+ \frac{f}{f'}\frac{v\gamma^2}{z^2}\left( z^2(h_{xy}+vh_{vy}) \right)' \nonumber\\
+ \frac{1}{z^3}\left(z^2(h_{vy}-z h_{vy}') \right)' = 0.
\end{align}
where $f$ is the metric function defined in \eqref{RNfdef}, and primes denote derivatives with respect to $z$.
Just like the QNM case, our holographic model UV completes the hydrodynamic analysis given above. In other words, we expect to find modes corresponding to the hydrodynamic poles we exhibited above, as well as an infinite tower of additional non-hydrodynamic modes which are specific to the holographic model we analyse. More specifically, in the transverse channel, we expect to find the hydrodynamic shear diffusion mode calculated in section \ref{sec:hydrotrans}, given by the dispersion relation
\begin{equation}\label{dispersion-transverse}
k = -i\frac{\epsilon + p}{\eta} v \cos\theta + O(k^2)\,,
\end{equation}
as well as a tower of higher non-hydrodynamic modes. In fact we will see a fascinating interplay of hydrodynamic poles and non-hydrodynamic poles, which gives rise to a rich analytic structure in the complex momentum plane recently observed in \cite{Withers:2018srf}. Such an analytic continuation suggests a path for a definition of QNMs in terms of hydrodynamic resummations, as indeed has been already explored in \cite{Heller:2015dha} for the Müller-Israel-Stewart theory \cite{muller1967paradoxon,israel1979transient}. The neutral case ($\mu = 0$) has been reported previously in \cite{Sonner:2017jcf} and so we are mostly interested in the behavior of these modes at finite chemical potential $\mu$.

\begin{figure}[h]
	\centering
	\begin{subfigure}{0.5\textwidth}
		\includegraphics[width=0.9\linewidth]{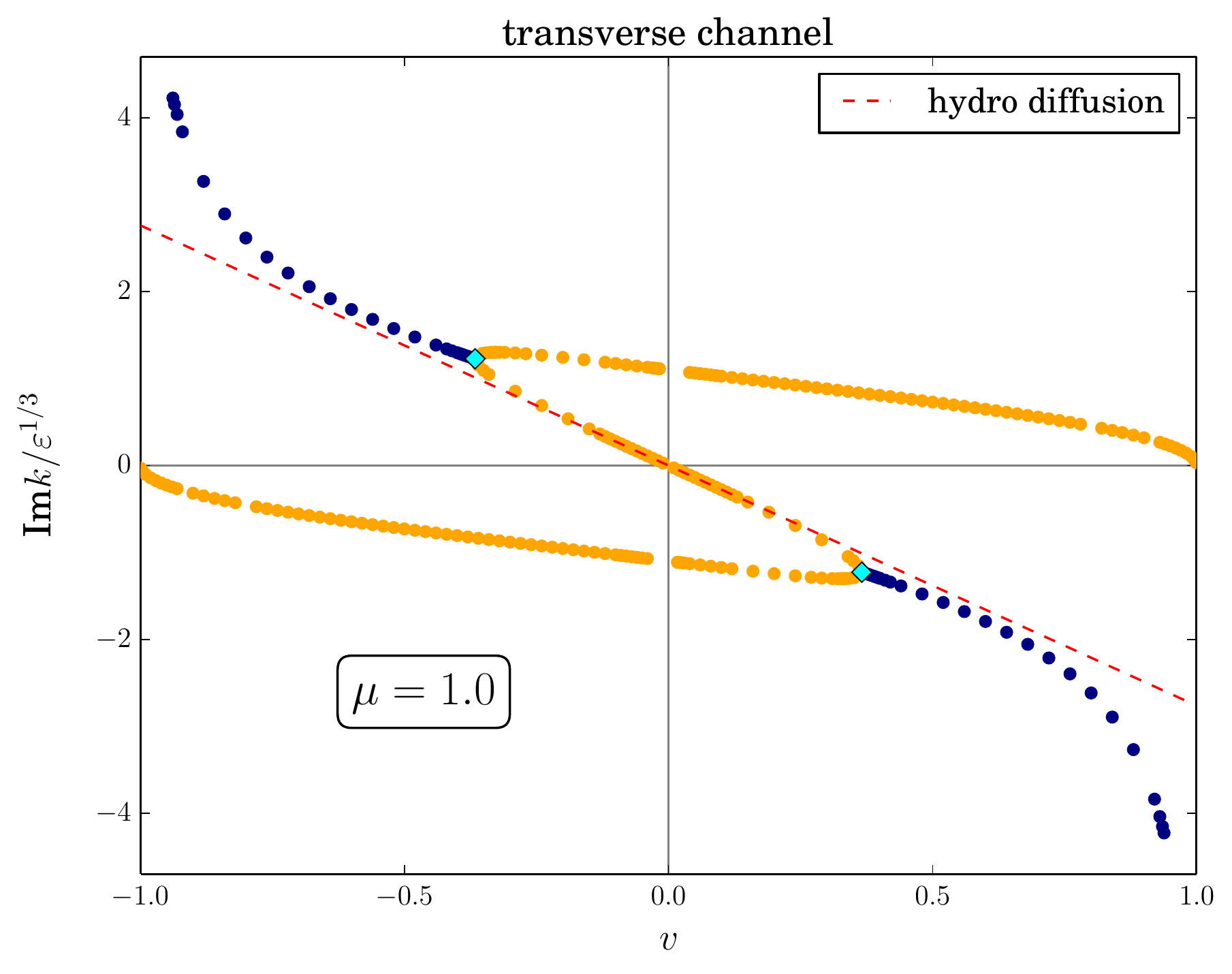}
	\end{subfigure}%
	\begin{subfigure}{0.5\textwidth}
		\includegraphics[width=0.9\linewidth]{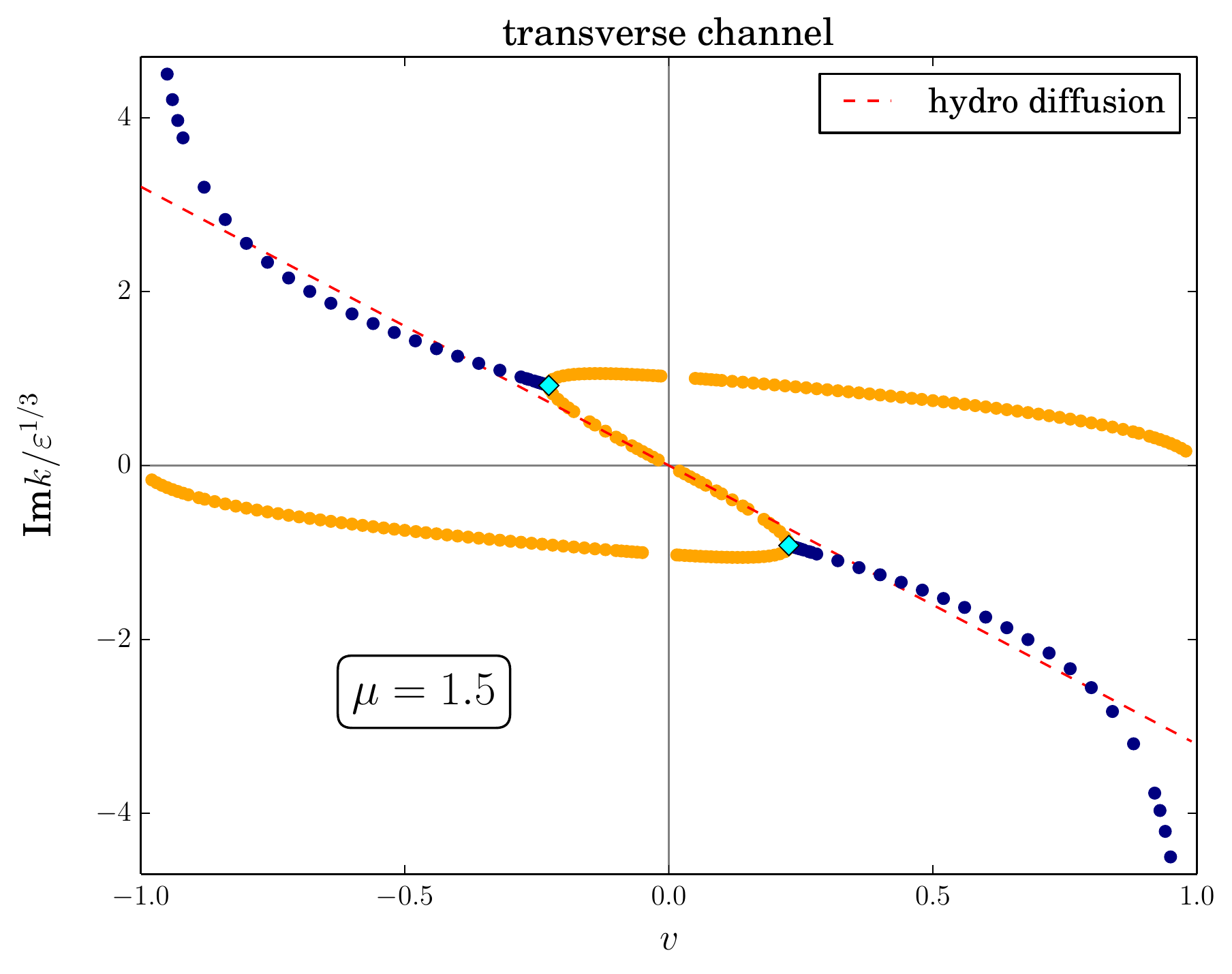}
	\end{subfigure}
	 \caption{(color figure) Spectrum of SCMs in the transverse channel for values of $\mu=1.0$ (left panel) and $\mu=1.5$ (right panel). The motion of the imaginary part of the complex momentum is shown as a function of background velocity. Parts of the spectrum where the mode is purely imaginary are shown in orange, while blue dots correspond to the parts where the mode has non-trivial imaginary and real parts. These behaviors transition into each other at the collision points indicated by turquoise diamonds. The red dashed line shows the SCM as predicted in our charged hydrodynamic effective theory.}
	 \label{fig:modes-transverse}
\end{figure}

In \autoref{fig:modes-transverse} we plot the spectrum of spatial collective modes, as a function of asymptotic flow velocity, for values of $\mu=1$ (left panel) and $\mu=1.5$ (right panel). We show the imaginary part of the complex momentum $k$ in suitable units along the vertical axis as a function of asymptotic flow velocity $v$ along the horizontal axis. The red dashed line represents the dispersion relation given by hydrodynamics (\ref{dispersion-transverse}). It can be seen that indeed one of the modes we find numerically corresponds to hydrodynamic shear diffusion in the low $k$ limit, while there exist, as expected, higher non-hydro poles. It is of particular interest to notice the pole collisions that take place between the hydrodynamical mode and a non-hydrodynamical mode for a critical value of the background flow velocity $v = v_{\rm critical}$ as indicated by turquoise diamonds in \autoref{fig:modes-transverse}: initially the poles have purely imaginary momentum $k$ (shown in orange), while, at the critical velocity $v_{\textrm{critical}}$, the poles collide with the resulting mode (shown in blue) acquiring a real part of $k$.

\begin{figure}[h]
	\centering
	\includegraphics[scale=0.5]{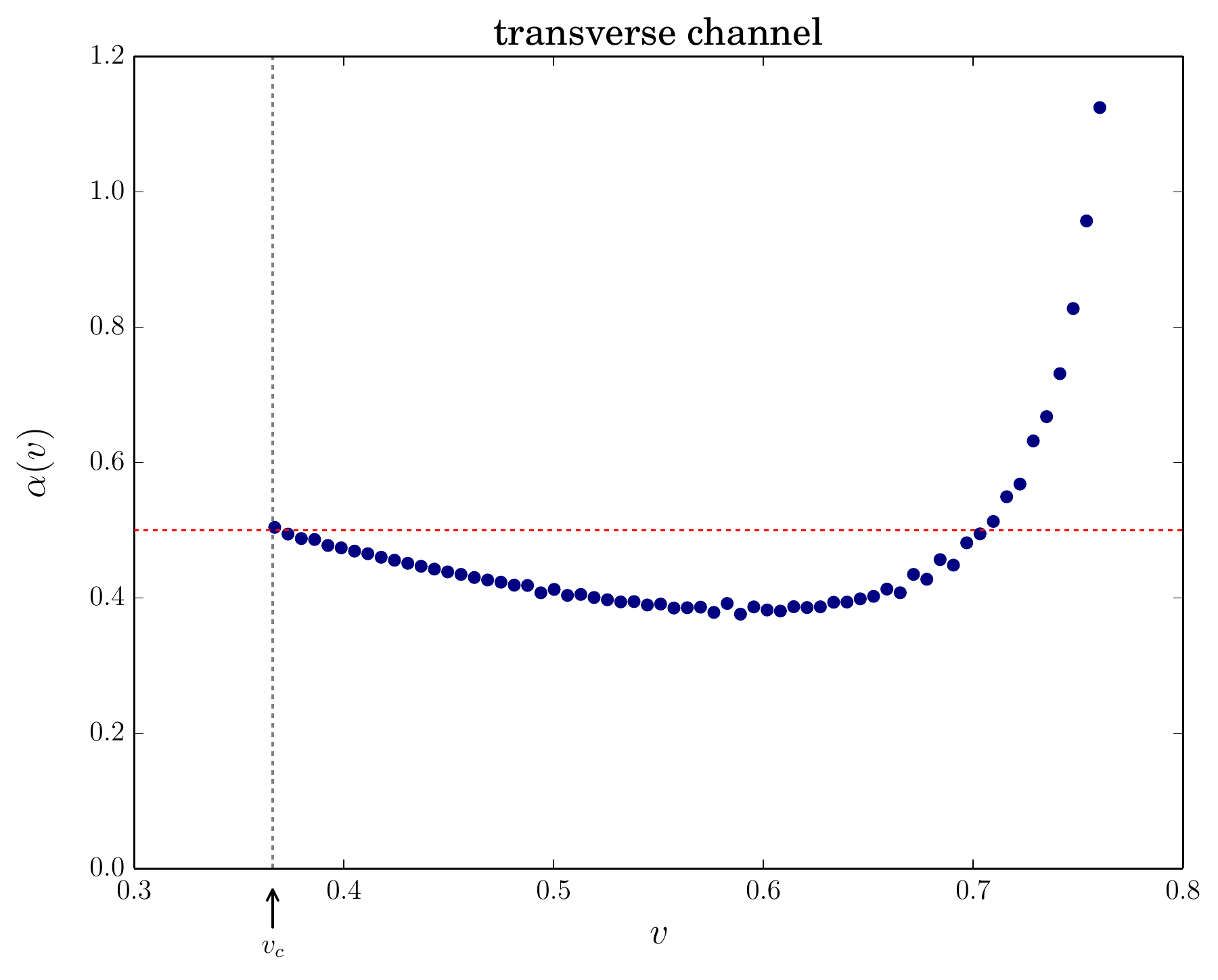}
	\caption{The critical exponent in the transverse channel for $\mu=1.0$, exhibited by the function $\alpha(v)$ defined in \eqref{eq.CriticalExponent}. It can be seen that the mean-field value of $\alpha = 1/2$ is approached at the critical velocity, $v_c$.}
	\label{fig:criticalexponenttrans}
\end{figure}

This pole collision is the SCM analog of similar QNM collisions, encountered previously in in \cite{Brattan:2010pq,Bhaseen:2012gg,Davison:2014lua,Withers:2016lft}.  The relevance to nonequilibrium phase transitions, as is also the case here, was first noted in \cite{Bhaseen:2012gg}.  Accordingly we find that the vanishing of the real part as one approaches the critical point follows a power law
\begin{equation}
\textrm{Re}\, k = A \, \left( v_{\textrm{critical}} - v \right)^{\alpha} \label{powerlawk}
\end{equation}
where $A$ is a proportionality coefficient, $v_{\textrm{critical}}$ is the critical velocity at which the transition occurs, and $\alpha$ is the critical exponent. We can numerically extract from this relation the value of the critical exponent $\alpha$. By defining $f(v) = \left(\textrm{Re}\, k \right)^2$ one can see that the following function
\begin{equation}\label{eq.CriticalExponent}
\alpha(v) = \frac{1}{2}\left[1-\frac{f''(v) \, f(v)}{\left(f'(v)\right)^2}\right]^{-1},
\end{equation}
where a prime denotes a derivative with respect to $v$, gives the critical exponent $\alpha$ in \eqref{powerlawk} when evaluated at the critical velocity, i.e. $\alpha = \alpha(v_c)$.
We use this relation to plot the behaviour of the critical exponent. By plotting the behaviour of this combination of derivatives for the range of velocities at which there exists a real part of the momentum (see \autoref{fig:criticalexponenttrans}), one can see that it converges to $\frac{1}{2}$ as we approach the critical velocity. This corresponds to the value of the critical exponent one might expect from a mean-field theory of this transition.

\begin{figure}[h]
	\centering
	\includegraphics[scale=0.5]{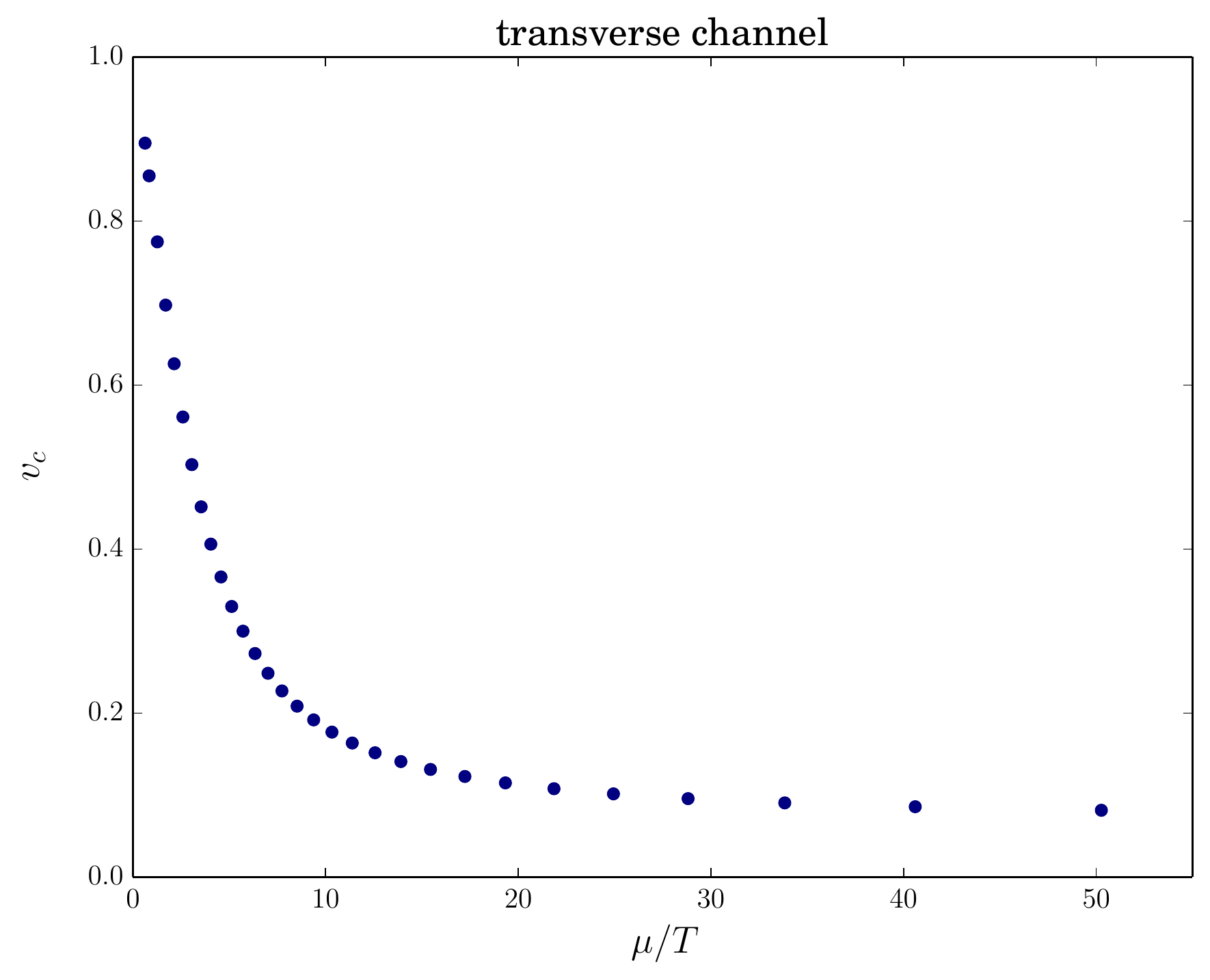}
	\caption{Behaviour of the critical velocity for flows of a holographic plasma at finite charge density as a function of $\mu/T$, above which the transverse channel SCM wavevector develops a real part, and the spatial relaxation becomes oscillatory. For specific values of $\mu/T$ these transitions are shown by the turquoise diamonds in \autoref{fig:modes-transverse}.}
	\label{fig:critical_velocity-transverse}
\end{figure}

We also analyzed how the critical velocity depends on the dimensionless ratio $\mu/T$. The results are shown in \autoref{fig:critical_velocity-transverse}. One can see that for very low $\mu/T$ (the high-temperature limit) the critical velocity approaches values nearing 1.

\subsubsection{Longitudinal channel}

The perturbation equations in the longitudinal channel are unwieldy and un-enlightening, so we do not give them here. We repeat a similar analysis as in the transverse case, mutatis mutandis.

\begin{figure}[h]
	\centering
	\begin{subfigure}{0.5\textwidth}
		\includegraphics[width=0.9\linewidth]{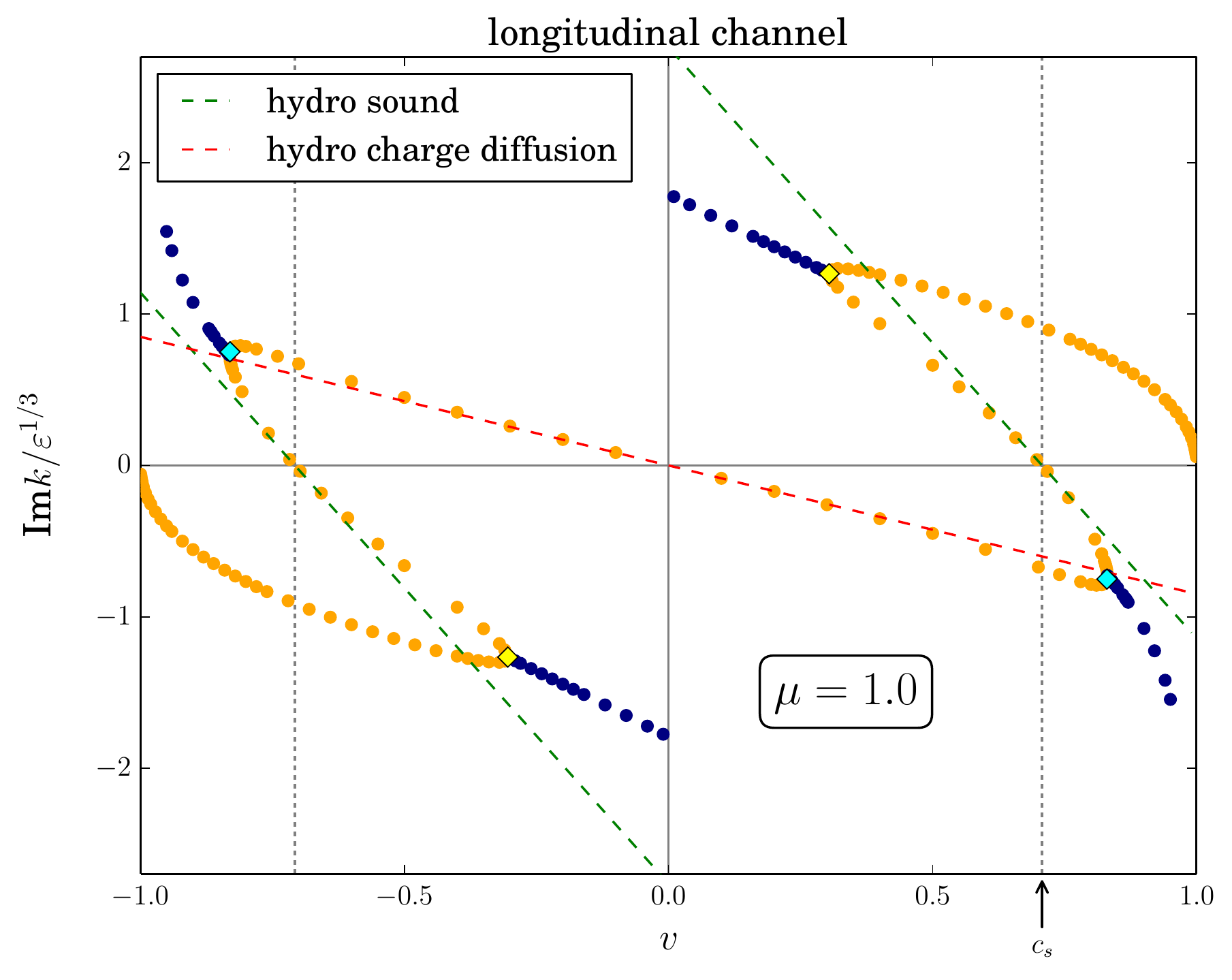}
	\end{subfigure}%
	\begin{subfigure}{0.5\textwidth}
		\includegraphics[width=0.9\linewidth]{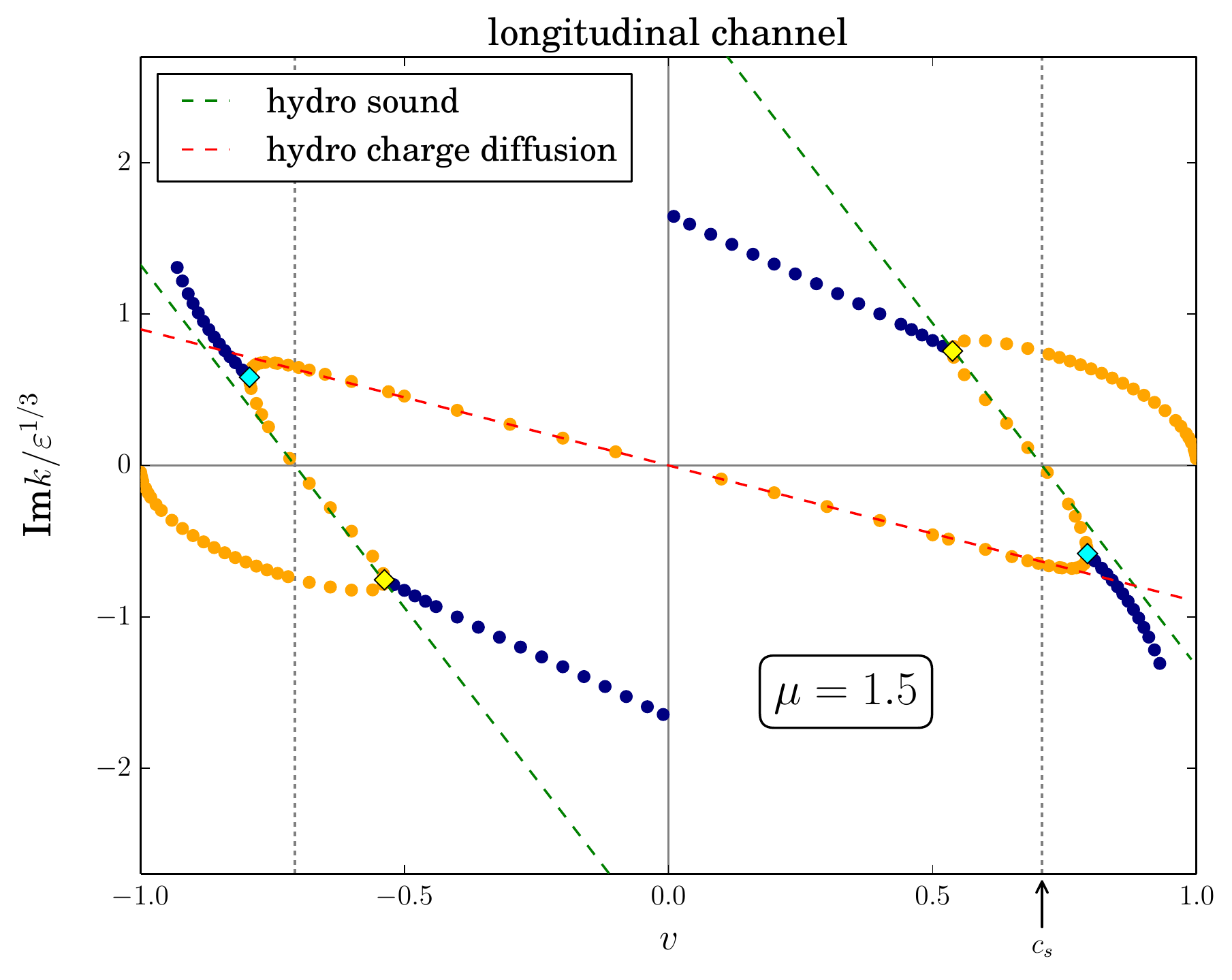}
	\end{subfigure}
	 \caption{(color figure) Spectrum of SCMs in the longitudinal channel for values of $\mu=1.0$ (left panel) and $\mu=1.5$ (right panel). As before purely imaginary-$k$ poles are shown in orange, while poles with both imaginary and real parts are shown in blue. In the longitudinal channel there are two kinds of pole collisions, where these behaviors transition into one another, one indicated by turquoise diamonds, and the other by yellow diamonds. Green and yellow lines show the predictions on the basis of our hydrodynamic effective theory.}
	 \label{fig:modes-longitudinal}
\end{figure}

From the hydrodynamic analysis in section \ref{sec:hydrolong}, in the longitudinal channel we expect to find hydrodynamic sound modes
\begin{equation}\label{sound}
k = -i\frac{2\cos \theta}{\Gamma (1-\sin^2\theta v^{2}_{0})^2\gamma_0}(v\mp v_0) + O(k^2),
\end{equation}
as well as a charge diffusion mode
\begin{equation}\label{chargediff}
k = -i \frac{\cos \theta}{D_{\alpha\beta}}v + O(k^2).
\end{equation}
The numerical spectrum of bulk SCMs in the longitudinal channel for $\mu=1.0$ and $\mu=1.5$ is given in \autoref{fig:modes-longitudinal}. The modes we calculate show good agreement with hydrodynamics (\ref{sound}) and (\ref{chargediff}) in the regime of small $k$, as required by hydrodynamics, and these are shown in \autoref{fig:modes-longitudinal} as the green and red dashed lines, respectively. As in the transverse case, we can again observe a pole collision at a certain critical velocity, indicated in \autoref{fig:modes-longitudinal} by yellow and turquoise diamonds. In the case indicated by the turquoise diamond the two sound and charge diffusion poles collide, after which the resulting pole has a non-zero value of Re $k$.  Both these poles are visible within a hydrodynamical analysis, and their point of collision appears to be well approximated by the extrapolation of their first-order in hydrodynamics dispersion relation as we discuss below. Within a purely hydrodynamical analysis, however, one does not observe the actual collision, the two modes remaining distinct through the would-be collision point. The yellow diamond corresponds to a collision between the hydrodynamic sound mode and a non-hydro SCM, which shows oscillatory behavior for small flow velocities, i.e. $v<v_c$, contrasting with the behavior of the collision between the two hydrodynamical modes marked in turquoise.

\begin{figure}[h]
	\centering
	\includegraphics[scale=0.5]{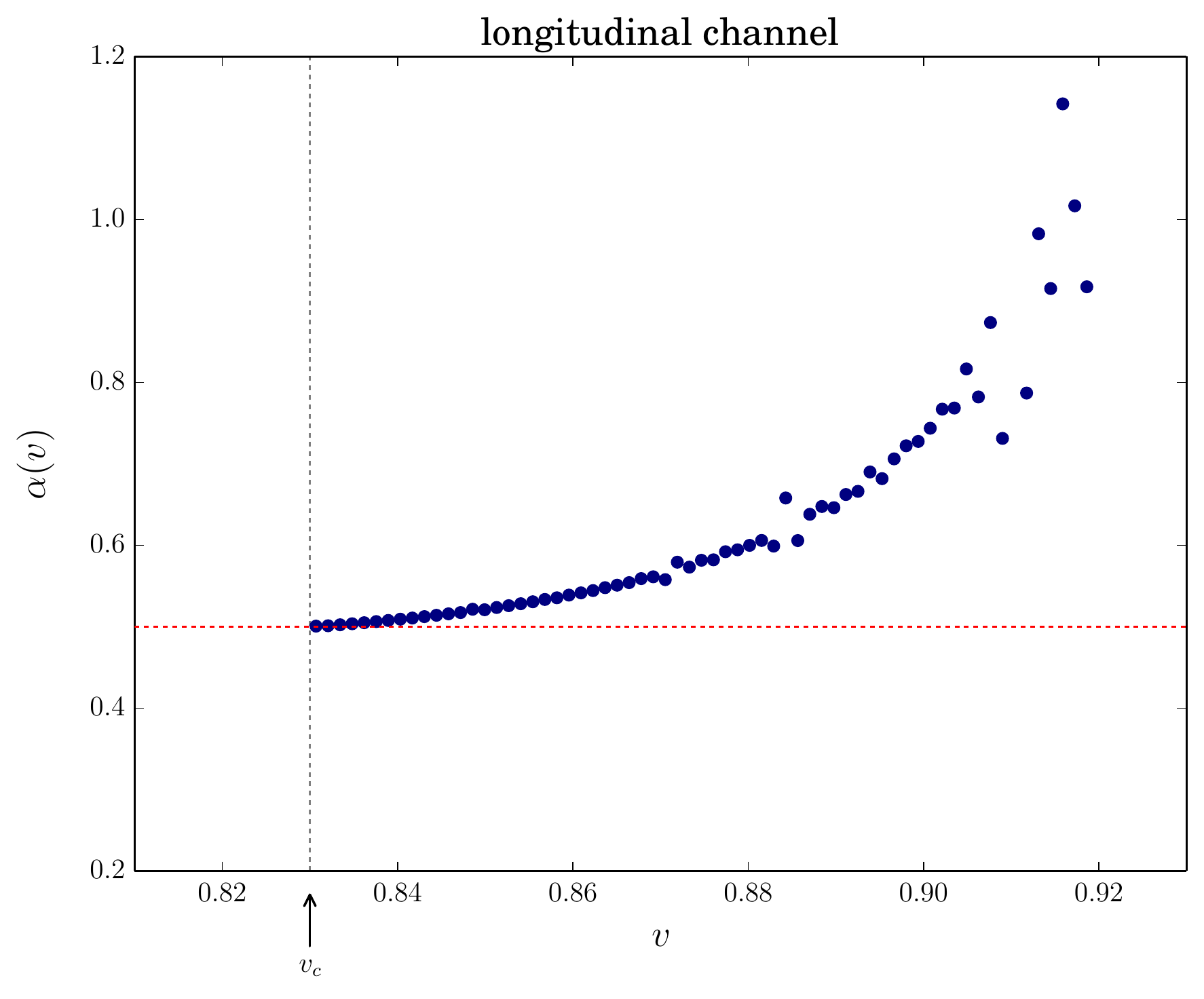}
	\caption{The critical exponent in the longitudinal channel for $\mu=1.0$, as exhibited by the function $\alpha(v)$ defined in \eqref{eq.CriticalExponent}. As for the analogous transition in the transverse channel \autoref{fig:criticalexponenttrans}, the value mean-field value $\alpha = 1/2$ is approached at the critical velocity, $v_c$.}
	\label{fig:criticalexponentlong}
\end{figure}

As shown in \autoref{fig:criticalexponentlong}, the value of the critical exponent again converges to $0.5$ as one approaches the critical velocity. This is in agreement with what we found in the transverse channel and, again, with expectations from mean-field theory. The numerical error in the computation of the expression \eqref{eq.CriticalExponent} becomes more pronounced as we go further away from the critical point, as evidenced by the noisier data. 

\begin{figure}[h]
	\centering
	\includegraphics[scale=0.5]{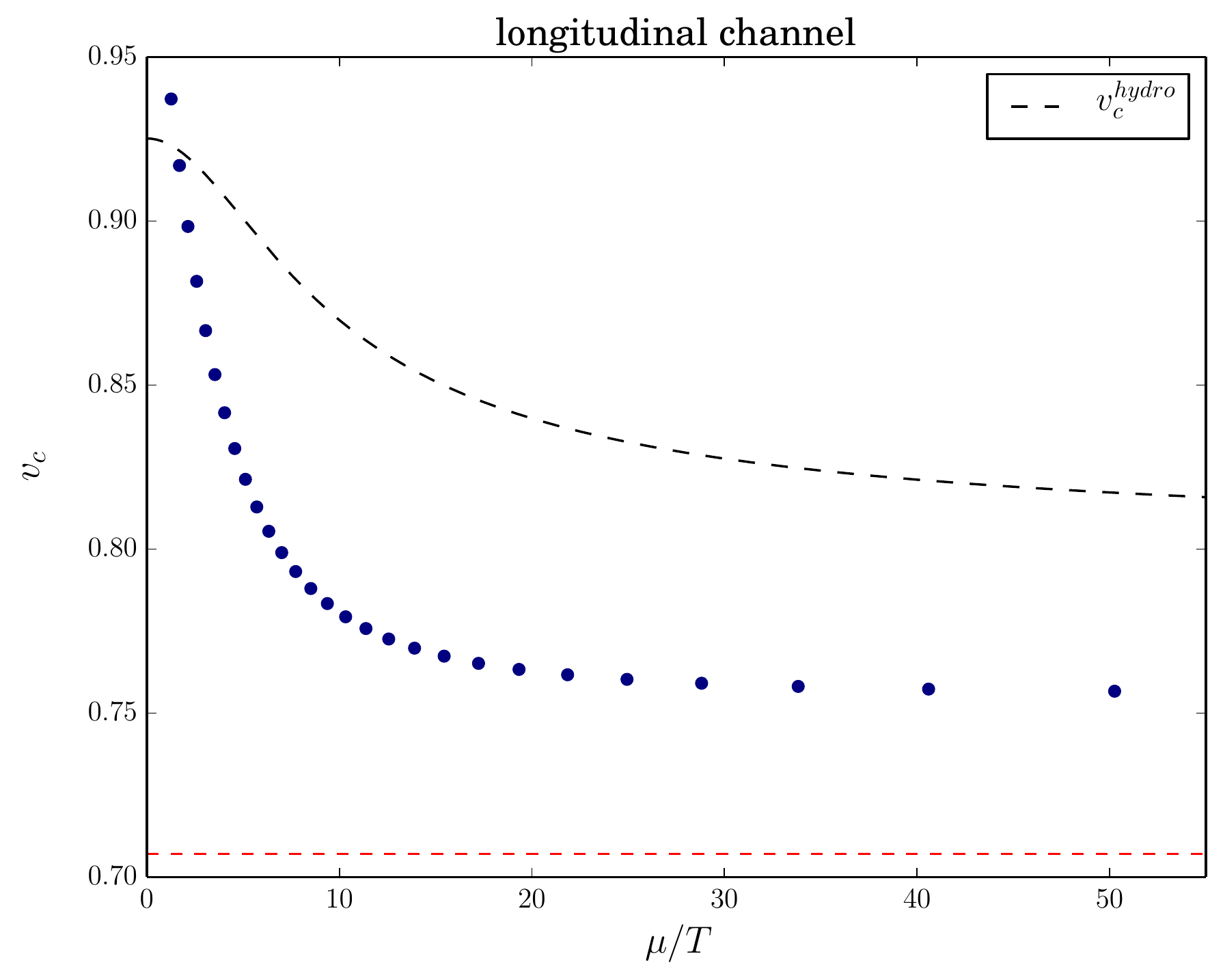}
	\caption{Behaviour of the critical velocity for flows of a holographic plasma at finite charge density as a function of $\mu/T$, above which the longitudinal channel SCM wavevector develops a real part and the spatial relaxation becomes oscillatory. For specific values of $\mu/T$ these transitions are shown by the turquoise diamonds in \autoref{fig:modes-longitudinal}. The dashed line indicates the hydrodynamic estimate for this collision point, $v_c^{\rm hydro.}$ given in \eqref{hydrocol}.}
	\label{fig:critical_velocity-longitudinal}
\end{figure}

Likewise, the dependence of the critical velocity on $\mu/T$ (\autoref{fig:critical_velocity-longitudinal}) is rather similar to the transverse case. In \autoref{fig:critical_velocity-longitudinal} we also show a hydrodynamic estimate for this collision obtained simply by equating the analytic expressions for $k(v)$ for the one of the sound modes \eqref{sound} with the diffusion mode \eqref{chargediff}, at normal incidence,
\be
v_c^{\rm hydro.} = \pm \frac{v_0}{1-\frac{\Gamma\gamma_0}{2D_{\alpha\beta}}}\label{hydrocol}.
\ee

\subsection{Spatial Collective Modes in a large number of dimensions}
In order to gain further analytic insight into the spectrum of SCMs in holographic models, we turn our attention to large-$d$ general relativity \cite{Asnin:2007rw, Emparan:2013moa,Emparan:2013xia}. In this approach one gains a small parameter, $1/d$, and an associated increase in analytic control. In this limit the set of quasinormal modes is partitioned by the scaling of their frequencies with $d$. In particular, there is one family of light modes with $\omega = \cO(d^0)$ and $q = \cO(d^{1/2})$, whose dispersion relations can be constructed analytically. These modes will be the focus of this section. 

The dispersion relations for the light modes in AdS were given in \cite{Emparan:2015rva} at charge neutrality, order-by-order in powers of $1/d$. In order to compute the corresponding SCM dispersion relations we simply perform a Lorentz boost to introduce a background velocity $v$, which transforms the frequency and wavenumber $(\omega,q)$ of \cite{Emparan:2015rva} to a perturbation with zero frequency and wavenumber $k(v)$ which we analytically continue into the complex plane. Restricting for simplicity to the case where $q,v,k$ are all in the $x$ direction, the Lorentz transformation relates them as follows,
\be
\omega = - \gamma v k,\quad  q = \gamma k. \label{largedboost}
\ee
From these relations and the large-$d$ scaling of $\omega,q$ we immediately see that we should treat $v = \cO(\omega/q) = \cO(d^{-1/2})$, which is in keeping with the scaling of the speed of sound for a conformal theory, $c_s = \cO(d^{-1/2})$. We further conclude that $\gamma = \cO(d^0)$ and thus $k = \cO(d^{1/2})$. Based on these scalings, let us define the order $d^0$ quantities:
\be
\bar{k} \equiv \frac{k}{\sqrt{d}},\qquad \bar{v} \equiv \sqrt{d} v, \label{largedbarred}
\ee
Our goal is to now find $\bar{k}(\bar{v})$ order-by-order in $d^{-1}$ from the dispersion relations given in \cite{Emparan:2015rva}. This can be achieved by replacing $\omega,q$ using \eqref{largedboost} then converting to barred quantities \eqref{largedbarred}. We then expand
\be
\bar{k}(\bar{v}) = \sum_{i=0}^\infty\frac{\bar{k}_i(\bar{v})}{d^i},
\ee
and solve for the coefficients $\bar{k}_i(\bar{v})$ in the $1/d$ Taylor expansion of the dispersion relation.

\subsubsection{Transverse channel}
The dispersion relation is given by \cite{Emparan:2015rva}\footnote{In the notation of \cite{Emparan:2015rva} $\bar{q}=\hat{k}_{\text{there}}$ and $d=D_{\text{there}}-1$.}
\be
\omega  = -i \bar{q}^2 \left(1 + \frac{1}{d^2}2\zeta(2) \bar{q}^2-\frac{1}{d^3}4\zeta(3)\left(\bar{q}^2+\bar{q}^4\right) + \frac{1}{d^4}8\zeta(4)\left(\bar{q}^2+7\bar{q}^4+\bar{q}^6\right) + O(d)^{-5}\right)\label{leading}
\ee
where we have introduced the order $d^0$ quantity, $\bar{q} \equiv q/\sqrt{d}$.
At leading order in the conversion to SCMs we find the following choice,
\be
\bar{k}_0(\bar{k}_0 + i \bar{v}) = 0. \label{largedshearroots}
\ee
Choosing $\bar{k}_0 = 0$ results in a trivial $\bar{k}$, and so this is a zero mode shifting the moduli of the equilibrium state. For the second root of \eqref{largedshearroots} we obtain a non-trivial mode, whose wavenumber $\bar{k}(\bar{v})$ can be written in a reasonably compact way by identifying an overall factor of $\gamma = \frac{1}{\sqrt{1-\bar{v}^2/d}}$,
\be
\frac{i\gamma}{\bar{v}}\bar{k} = 1+2\bar{v}^2\frac{\zeta(2)}{d^2}-4\bar{v}^2(1-\bar{v}^2)\frac{\zeta(3)}{d^3} + 2\bar{v}^2(4-13\bar{v}^2+4\bar{v}^4)\frac{\zeta(4) }{d^4} + \cO(d^{-5}).
\ee
From this expression we conclude that Re$\bar{k} = \cO(d^{-5})$.

\subsubsection{Longitudinal channel}
The dispersion relation is given by \cite{Emparan:2015rva}
\bea
\omega =& -i \bar{q}^2\bigg[&1-\frac{1}{d}-\frac{1}{d^2}\left(1-\frac{\pi^2}{3}\bar{q}^2\right) - \frac{1}{d^3}\left(1+\left(\frac{4\pi^2}{3} + 8\zeta(3)\right)\bar{q}^2 + 4\zeta(3)\bar{q}^4\right)\nonumber\\
&&-\frac{1}{d^4}\left(1+\left(\frac{\pi^2}{3}-\frac{\pi^4}{9}- 16\zeta(3)\right)\bar{q}^2-\left(\frac{31\pi^4}{45}+36\zeta(3)\right)\bar{q}^4 - \frac{4\pi^4}{45}\bar{q}^6\right)\bigg]\nonumber\\
&\pm \bar{q}\bigg[& 1+\frac{1}{d}\left(\frac{1}{2}+\bar{q}^2\right)+\frac{1}{d^2}\left(\frac{3}{8}+\left(\frac{\pi^2}{3} -\frac{1}{2}\right)\bar{q}^2-\frac{1}{2}\bar{q}^4\right)\nonumber\\
&&+\frac{1}{d^3}\left(\frac{5}{16} - \left(\frac{9}{8}+\frac{\pi^2}{6}+4\zeta(3)\right)\bar{q}^2 + \left(\frac{3}{4}+\pi^2 - 2\zeta(3)\right)\bar{q}^4 + \frac{1}{2}\bar{q}^6\right)\nonumber\\
&&+\frac{1}{d^4}\Bigg(\frac{35}{128}-\left(\frac{25}{16}+\frac{3\pi^2}{8}-\frac{4\pi^4}{45}-2\zeta(3)\right)\bar{q}^2 + \left(\frac{13}{16}-\frac{3\pi^2}{2}+\frac{29\pi^4}{45}-5\zeta(3)\right)\bar{q}^4\nonumber\\
&&-\left(\frac{5}{4}+\frac{5\pi^2}{6}-\frac{\pi^4}{15}+22\zeta(3)\right)\bar{q}^6 -\frac{5}{8}\bar{q}^8\Bigg)\bigg]+\cO(d^{-5})
\eea
Similarly at order $d^0$ we find
\be
i\bar{k}_0\left(\bar{k}_0+i (\bar{v}\pm1)\right) = 0
\ee
The choice $\bar{k}_0$ leads to a zero-mode, whilst the other root leads to,
\bea
i\gamma\bar{k} &=& (\bar{v}\pm 1) 
+ \frac{1}{d}\Bigg[-\bar{v} \pm\left(\frac{1}{2} - \bar{v}^2\right)\Bigg] 
+ \frac{1}{d^2}\Bigg[\pm\left(\frac{7}{8}+ \frac{\bar{v}^2}{2}  - \frac{\bar{v}^4}{2} \right) +  \left( 2\bar{v} \pm 4\bar{v}^2 +2\bar{v}^3\right)\zeta(2)\Bigg]\nonumber\\
&&+\frac{1}{d^3}\Bigg[\pm\left(\frac{25}{16} - \frac{3\bar{v}^2}{8} + \frac{3\bar{v}^4}{4} - \frac{\bar{v}^6}{2}\right) 
+ \left(2\bar{v} -8\bar{v}^3  \mp 2\left(\bar{v}^2 +2\bar{v}^4\right)\right)\zeta(2) \nonumber\\
&&+ \left(-4\bar{v} + 24\bar{v}^3 + 4\bar{v}^5 \pm \left(-2 + 8 \bar{v}^2 + 18 \bar{v}^4\right)\right)\zeta(3)\Bigg]\nonumber\\
&&+\frac{1}{d^4}\Bigg[\pm\left(\frac{363}{128}-\frac{11\bar{v}^2}{16} - \frac{7\bar{v}^4}{16}+ \frac{5\bar{v}^6}{4} - \frac{5\bar{v}^8}{8}\right)
+\left(4\bar{v} + 2\bar{v}^3 \pm \left(\frac{3\bar{v}^2}{2} + 6\bar{v}^4 - 2\bar{v}^6\right)\right)\zeta(2)\nonumber\\
&&+\left(-4\bar{v} - 16\bar{v}^3-36\bar{v}^5 \pm \left(-5+10\bar{v}^2 -55\bar{v}^4 - 6 \bar{v}^6\right)\right)\zeta(3)\nonumber\\
&&+\left(-34\bar{v} - 98\bar{v}^3 + 100\bar{v}^5 + 8\bar{v}^7 \pm\left(-112\bar{v}^2 + 38\bar{v}^4 + 50 \bar{v}^6\right)\right)\zeta(4)\Bigg] + O(d)^{-5}.
\eea
Once more, from this expression we conclude that Re$\bar{k} = \cO(d^{-5})$. We expect that Re$\bar{k} = 0$ continues to all perturbative orders in $1/d$. 

\subsection{BTZ black hole}\label{sec.BTZSCM}
An example in which we can find the exact SCM analytically \cite{Sonner:2017jcf} and without any approximation is the three-dimensional BTZ black hole, dual to the thermal ensemble of a two-dimensional relativistic CFT. We follow the conventions of \cite{Son:2002sd}. In particular we will study modes of a probe scalar field, $\phi(t,\rho,\varphi) \sim \Phi(\rho)e^{-i\omega t + i q \varphi}$, of mass $m$ in the background
\be
ds^2 =  - \frac{\rho^2 - \rho_+^2}{\ell^2 }dt^2 + \rho^2  d\varphi^2 + \frac{\ell^2}{\rho^2 - \rho_+^2}d\rho^2.
\ee
This background has Hawking temperatures $T = \frac{\rho_+}{2\pi\ell^2}$ which are related to its mass $M$ as
\be
M = \frac{\pi^2 \ell^2}{4 G_N} T^2.
\ee
The strategy here is to exploit the Lorentz invariance of the underlying CFT$_2$ to construct the SCM from the known spectrum of quasinormal modes \cite{Cardoso:2001hn,PhysRevD.64.064024,Son:2002sd}. For a scalar field $\phi$, dual to an operator $\cO$ of dimension $\Delta = 1+\sqrt{1 + m^2 \ell^2}$, the quasinormal modes occur in two integer series, \cite{Son:2002sd}
\be\label{eq.QNMSeriesBTZ}
\omega_n^{(\pm)} = \pm q - 4\pi i T \left( \frac{\Delta}{2} + n \right)\,,\qquad n \in \mathbb{Z}^{*}.
\ee

\begin{figure}[h]
\begin{center}
\includegraphics[width=0.5\textwidth]{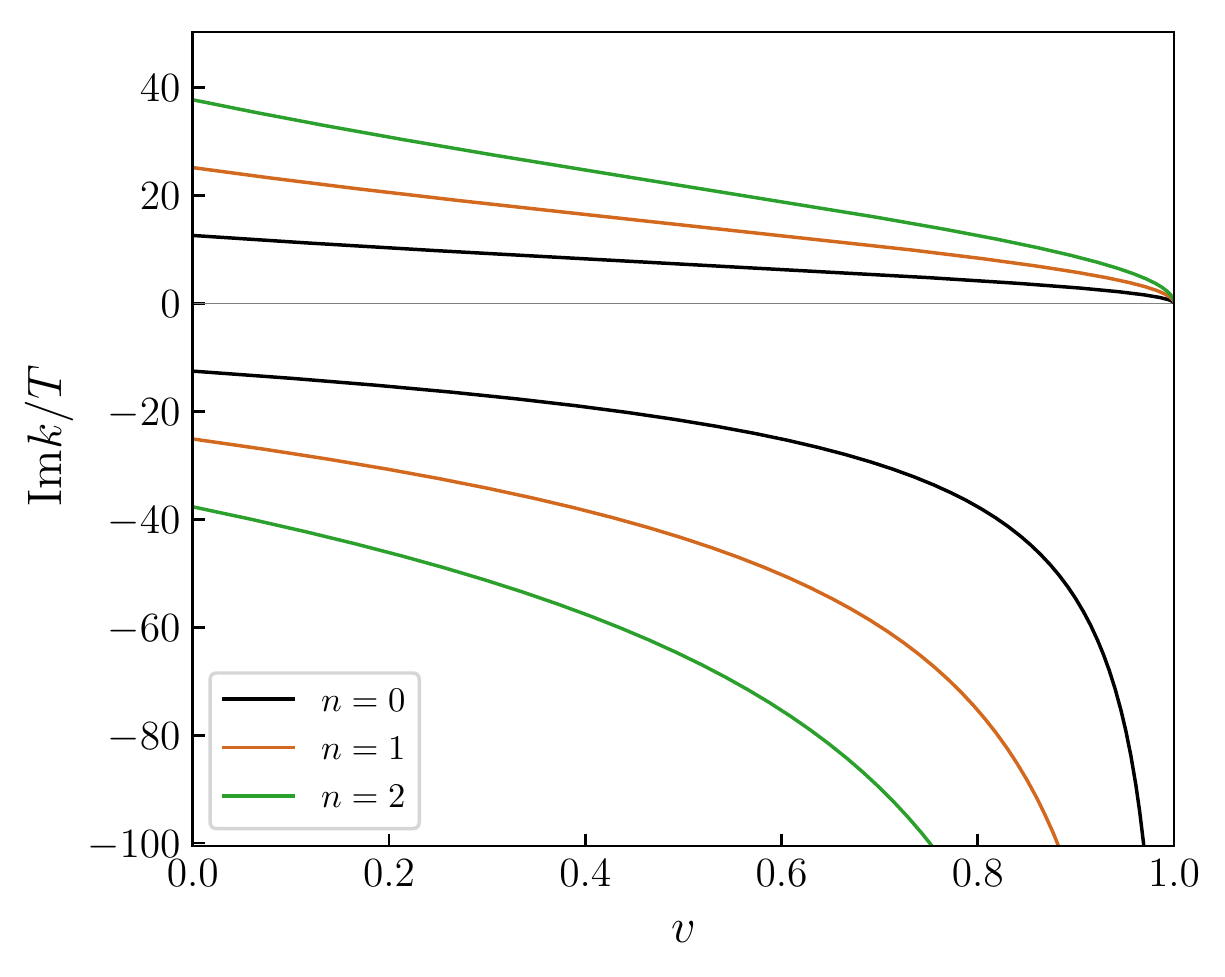}
\caption{(color figure) First three modes of the spectrum of SCM of a neutral CFT$_2$ at temperature $T$, as deduced from the mode spectrum of the boosted BTZ solution. The positive branches correspond to $k_+$ modes, while the negative branches correspond to $k_-$ modes. \label{fig.btz}}
\end{center}
\end{figure}

We are, as before, interested in purely spatial modes at complex momentum in a frame where the fluid has a finite background velocity. The four-momentum of such a mode is thus given by $k_{\rm lab}^\mu = (0, k)^\mu$ in the lab frame. We can then boost to the fluid rest frame to obtain
\be\label{eq.boostedOmegaK}
\omega = -\gamma v k\,, \qquad q= \gamma k\,.
\ee
Plugging these values into equation \eqref{eq.QNMSeriesBTZ} and solving for $k$, we find two series of purely imaginary SCM at
\be\label{eq.BTZSCM}
k_\pm = i \frac{4\pi T}{\gamma \left( v \pm 1 \right)} \left( \frac{\Delta}{2} + n \right)  \,.
\ee
We show the resulting SCM spectrum in \autoref{fig.btz}.

Let us note that in this case there is an intuitive explanation in terms of the relativistic Doppler shift. The prefactor on the right-hand side of \eqref{eq.BTZSCM}, namely $\frac{T}{\gamma \left( v \pm 1 \right)}$, can be interpreted as the Doppler shifted temperature
\be
T^{\rm Doppler} = \sqrt{\frac{v\pm 1}{v\mp 1}}\,T
\ee
experienced by an observer in relative motion measuring a black-body spectrum, or equivalently the black-body spectrum with respect to a Doppler shifted wavelength if the observer instead chooses to interpret the spectrum keeping the temperature constant. This simple relationship between the quasinormal modes and SCM is special to the BTZ case. As one can easily verify, it depends crucially on the fact that $\omega$ depends only linearly on momentum, while any non-linear correction term would spoil this simple relationship. We have explicitly confirmed that our higher-dimensional SCM do not follow a functional form that is simply given by a Doppler-shifted temperature. Finally, it is clear that in non-relativistic systems no relationship relying on boost symmetry will be applicable.

Somewhat remarkably, in three dimensions we can actually go further.

\section{Black Janus: a fully non-linear analytic example}\label{sec.BlackJanus}
As we have argued, SCMs describe steady states. But, up to this point, the evidence we presented was numerical \cite{Sonner:2017jcf} (see section \ref{sec.NonlinearNumerics} below for more details on the numerical method employed in that work), due to the difficulty in obtaining analytic solutions describing the requisite non-linear steady states.
In fact, analytical backgrounds illustrating the importance of SCM can be obtained. A particularly sharp example of this is provided by the finite $T$ black Janus solution of \cite{Bak:2011ga}.
This solution describes a finite $T$ defect solution created by turning on a step function for a source of a scalar operator in the field theory. As argued previously, the physics of the SCM is universal and therefore independent of how they are excited, and here that excitation is the spatial variation of the scalar source.
In fact, we now show that the black Janus solution may globally be regarded as a self-consistently backreacted sum of infinitely many SCMs. This makes it clear that at large distances from the obstacle, as predicted from our analysis, a single SCM dominates the spatial profile. We will now demonstrate this structure explicitly.

Having this analytic solution will also allow us to take the $T=0$ limit, where the SCMs coalesce in the complex $k$-plane and the Janus tail becomes power-law. This emergence of a branch cut from the coalescence of poles is expected to be relevant also in other examples. 

The black hole of \cite{Bak:2011ga} at temperature $T$ may be written in the following form
\bea
ds^2 &=& f(\mu)^2(-\sinh^2(2\pi T p)\, d\tau^2 + dp^2+d\mu^2), \\
\phi &=& \phi(\mu)
\eea
where
\bea
f(\mu)^{-1} &=& \frac{\sqrt{1+k}}{2\pi T}\,{\sf cd}\left(\frac{2\pi T\mu}{\sqrt{1+{\sf m}}},{\sf m}\right),\\
e^{\frac{1}{2}\phi} &=& \frac{1+ \sqrt{k}\,{\sf sn}\left(\frac{2\pi T\mu}{\sqrt{1+{\sf m}}},{\sf m}\right)}{{\sf dn}\left(\frac{2\pi T\mu}{\sqrt{1+{\sf m}}},{\sf m}\right)}\,,
\eea
where ${\sf sn}$, ${\sf cd}$ and ${\sf dn}$ are Jacobi elliptic functions and ${\sf m}$ is the elliptic modulus\footnote{This is denoted `$k$' in the original references, but we prefer to relabel is `${\sf m}$' in order to avoid any confusion with the momentum $k$.}. 

\begin{figure}[h]
\begin{center}
\includegraphics[width=0.7\textwidth]{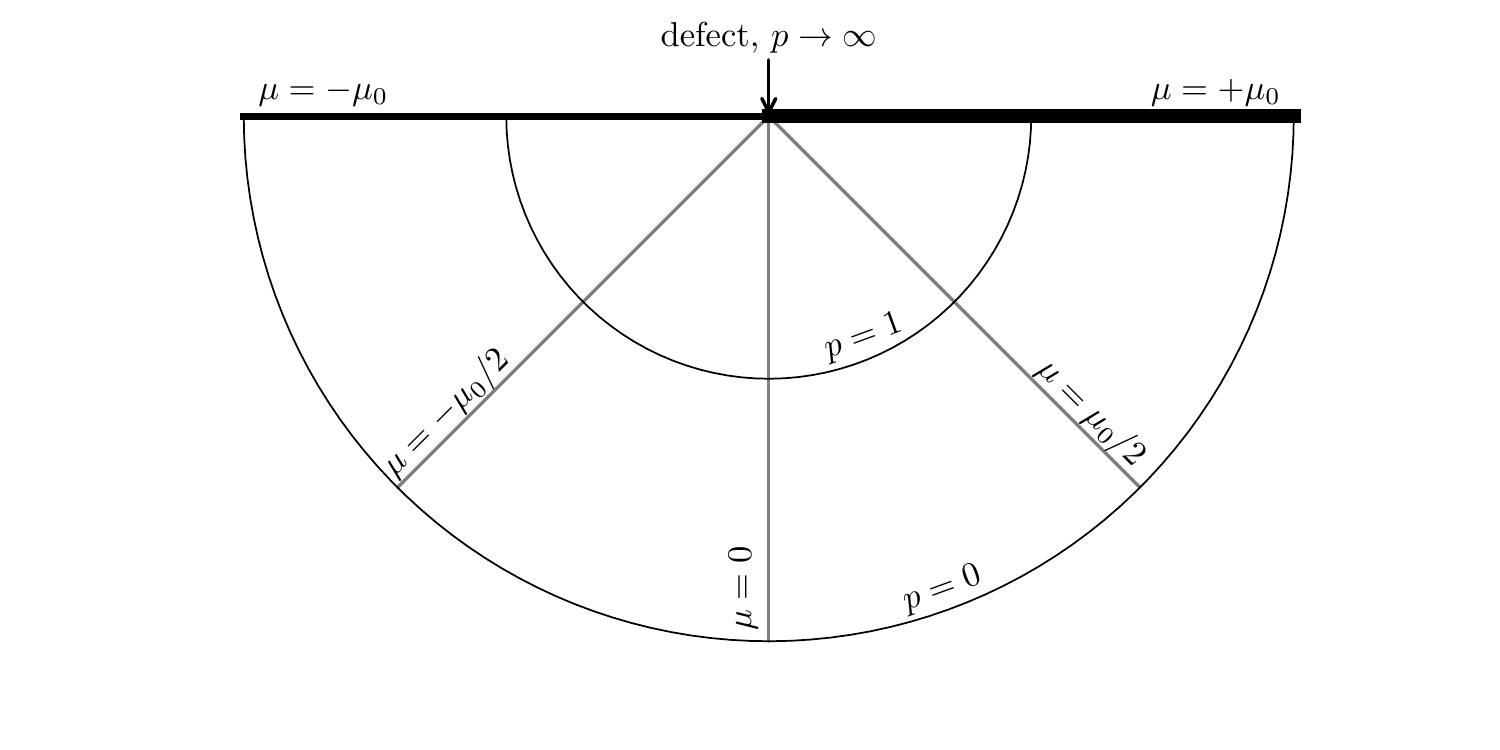}
\caption{Coordinates for the black Janus solution. The defect is located at the origin $p=0$. We reach the asymptotic regions on either side by taking $p\rightarrow\infty$ with angular coordinate $\mu = \pm \mu_0$ \label{fig.blackJanus}. In the asymptotic regions the scalar field approaches the limiting value $\phi(\pm \mu_0) =  \pm2 \log\left(\frac{1+ \sqrt{{\sf m}}}{\sqrt{1-{\sf m}}}\right)$.} 
\end{center}
\end{figure}

This is a solution to the Einstein-massless scalar equations of motion for $d=2$ with negative cosmological constant \cite{Bak:2011ga}.
It is described by an angular coordinate $\mu$ and a radial coordinate $p$. The angle is bounded by $ |\mu| \leq \mu_0 \equiv \frac{\sqrt{1+{\sf m}}}{2\pi T}K({\sf m})$ where $K$ is the complete elliptic integral of the first kind, the limit being saturated on the AdS boundary. The defect itself can be reached by taking $p\to\infty$ at fixed $\mu$. An illustration of the geometry and the coordinates employed is given in \autoref{fig.blackJanus}.

Finally, the solution is parameterised by the elliptic modulus ${\sf m}$, which is the parameter that determines the size of the step, i.e. it dictates the value of the scalar field source on either side of the defect, through the relation
\be
\phi(\pm \mu_0) =  \pm2 \log\left(\frac{1+ \sqrt{{\sf m}}}{\sqrt{1-{\sf m}}}\right).
\ee

\subsection{One-point functions}
We first focus our attention on the CFT one-point functions, which we compute by going to Fefferman-Graham coordinates, $(t,x,\rho)$, near the boundary on either side of the defect, i.e $\mu = \pm\mu_0$. In the new coordinates the boundary is given by $\rho = 0$, and the coordinate transformation is given by
\bea
\tau &=& t + O(\rho)^4\nonumber\\
\mu &=& \pm\mu_0 + \csch(2\pi Tx)\,\rho - \frac{1}{24}(2\pi T)^2\csch^3(2\pi Tx)(5+3\cosh(4\pi Tx))\,\rho^3+ O(\rho)^4\nonumber\\
p &=& -\frac{1}{\pi T}\,\log\left(\pm\tanh(\pi T x)\right) - \pi T \coth(2\pi Tx)\csch(2\pi Tx)\,\rho^2 + O(\rho)^4
\eea
Note that we take $x>0$ for all $0\leq p < \infty$ at $\mu = \mu_0$ and similarly $x<0$ for all $0\leq p < \infty$ at $\mu = -\mu_0$. We do not consider the defect location $x=0$ directly. The metric and scalar field in these coordinates are given by the expressions,
\bea
ds^2 &=& \frac{d\rho^2}{\rho^2}- \left(\frac{1}{\rho^2}-\frac{(2\pi T)^2}{2}\right)dt^2 + \left(\frac{1}{\rho^2}+\frac{(2\pi T)^2}{2}\right)dx^2 +O(\rho)^2,\\
\phi &=& \pm\left(2 \log\left(\frac{1+ \sqrt{{\sf m}}}{\sqrt{1-{\sf m}}}\right) - \frac{\sqrt{k} (2\pi T)^2\csch^2(2\pi Tx)}{1+k} \rho^2+ O(\rho)^4\right).
\eea
where the upper sign corresponds to the right asymptotic region and the lower sign to the left asymptotic region.
There is no $\rho^2 \log\rho$ term since the source is constant on either side of the defect. Going into the bulk, the metric eventually becomes deformed by the scalar backreaction at order $\rho^2$ as we outline below. 
From the above results, using holographic renormalisation \cite{deHaro:2000vlm}, we find the expectation values
\bea
\left<T_{ab}\right> &=& (2\pi T)^2 \delta_{ab}\\
\left<O_\phi\right> &=& \mp2\frac{\sqrt{{\sf m}}(2\pi T)^2}{1+{\sf m}} \csch^2(2\pi T x). \label{Janusvev}
\eea

\subsection{Extracting the SCM}
We begin our comparison by studying the behaviour of the one-point functions. For this we recall that the spectrum of SCM for a scalar field on BTZ were computed in \eqref{eq.BTZSCM}. In the present analysis we only need the $v=0$ cases. Specializing to a massless field, we have $\Delta =2$, whence
\be
k_n^{\pm} = \pm i 4\pi T(1+n).\label{JanusSCMs}
\ee
Then, we can explicitly see that \eqref{Janusvev} can be expressed as a linear sum of these modes, on either side of the defect. On the $x>0$ side of the defect we have
\be\label{eq.JanusSCMRight}
\left<O_\phi\right> =  \sum_{n=0}^{\infty}A^+_ne^{i k^+_n x}\qquad \textrm{with}\qquad A^+_n = -\frac{8\sqrt{{\sf m}}(2\pi T)^2}{1+{\sf m}}(n+1),
\ee
and on the $x<0$ side of the defect we have,
\be
\left<O_\phi\right> =  \sum_{n=0}^{\infty}A^-_ne^{i k^-_n x}\qquad \textrm{with}\qquad A^-_n = \frac{8\sqrt{{\sf m}}(2\pi T)^2}{1+{\sf m}}(n+1).
\ee
For large distances from the obstacle one readily sees that the solution falls off exponentially in either direction with a characteristic length scale ${k_0^-} = 4\pi T$ to the left and $|k_0^+| = 4\pi T$ to the right, in other words according to the dominant SCM in either sector.

The above one-point functions combined with metric and scalar boundary conditions are sufficient to determine the full bulk solution.
Thus, the one-point functions tell us that the bulk black Janus solution is simply the result of summing infinitely many SCMs and matching at a defect boundary condition.
It may be tempting to then conclude that the full bulk metric is a linear sum of SCMs, in the way that the one-point functions are. However this is only true asymptotically, near the AdS boundary. Specifically, up to and including stress-tensor and expectation-value order in the small $\rho$ expansion there is no trace of backreaction of the modes. However, moving to higher powers in $\rho$ one encounters non-linear backreaction, (presented here for the $x>0$ side)
\be
\delta (ds^2) = -\left(\frac{A_0^+}{16}\right)^2 \csch^4\left(2\pi T x\right) (dt^2+dx^2)\rho^2 + O(\rho)^4\,,
\ee
where $\delta (ds^2)$ is the difference between the linearized and the fully backreacted metric, e.g. $\delta(ds^2) = ds^2 -\left(ds^2\right)_{\rm lin}$. Subsequently, at even higher powers of $\rho$, the scalar modes can interact with each other resulting in non-linear adjustments to the $\phi$ profile, 
\be
\delta\phi =  -\frac{2}{3}\left(\frac{A_0^+}{16}\right)^3 \csch^6\left(2\pi T x\right)\rho^6 + O(\rho)^8\,,
\ee
with the $\delta$ notation defined as an obvious extension of the above.
Thus we regard the black Janus solution as a self-consistently backreacted solution resulting from the linear sum of infinitely many BTZ SCMs. 

\subsubsection{Zero-temperature limit and analytical structure}
Finally, we would like to comment on the $T=0$ limit. Here the expectation value \eqref{Janusvev} becomes power-law in $x$ rather than exponential,
\be\label{eq.PowerLawDecay}
\left<O_\phi\right> = \mp2\frac{\sqrt{{\sf m}}}{1+{\sf m}} \frac{1}{x^2}\,, \qquad \qquad (T=0),
\ee
and in this limit the SCMs appear to accumulate at the origin \eqref{JanusSCMs}, which becomes in fact a branch point,  conforming with the usual intuition of power laws coming from branch cuts. 

\begin{figure}[h]
\begin{center}
\includegraphics[width=0.8\textwidth]{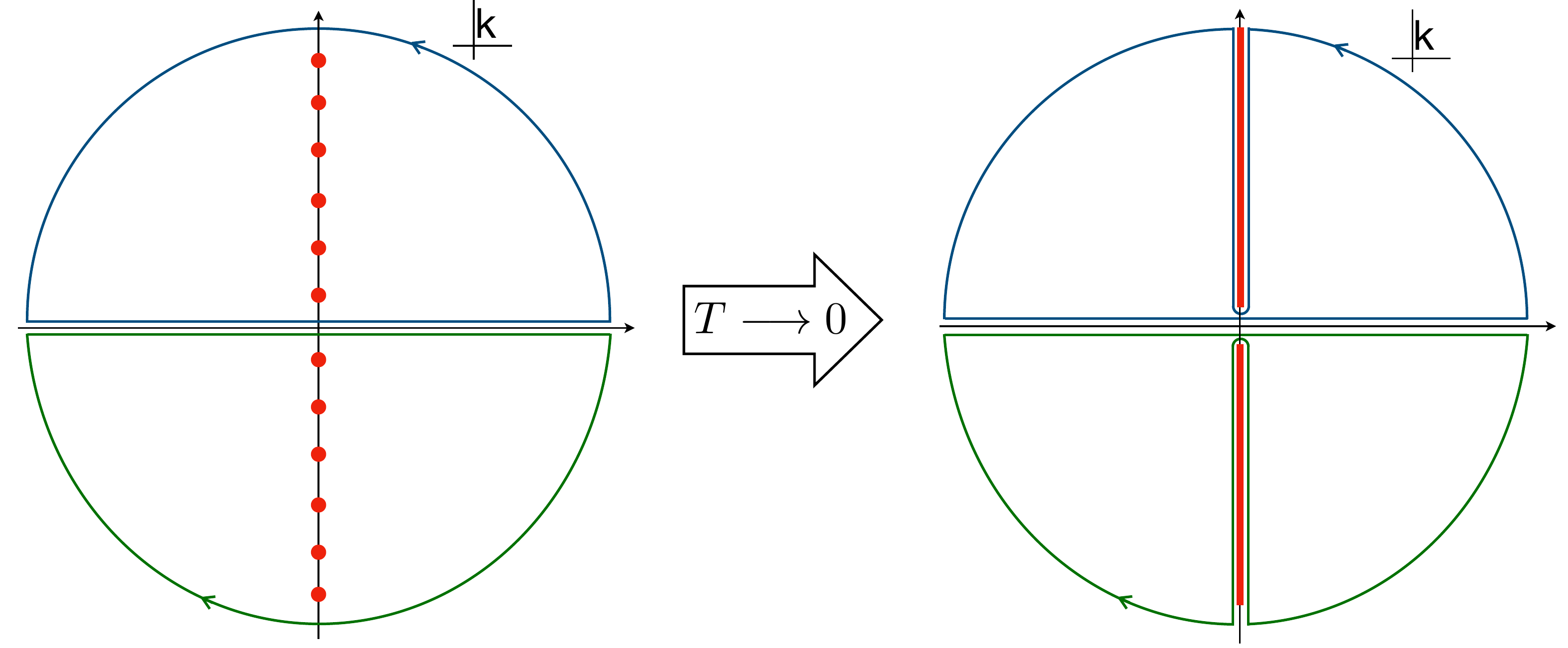}
\caption{(color figure) Analytic structure in the complex momentum plane for the black Janus solution. At finite temperature we have two sectors of SCM along the positive and negative imaginary axis (we summarize all poles, that is both those of $G^{\scriptscriptstyle [\searrow]}(k)$ and $G^{\scriptscriptstyle [\swarrow]}(k)$ in the same figure). In the $T\rightarrow 0$ limit these condense into a continuum in the form of two branch cuts along positive and negative imaginary axes. This continuum, for example, manifests itself as a power-law decay at large distances \eqref{eq.PowerLawDecay} or as a discontinuity in the spectral representation of the two point function. \label{fig.polesToBranch}} 
\end{center}
\end{figure}

We can make this precise by writing the expectation value as a Laplace transform
\be\label{eq.OphiLaplace}
\left<O_\phi\right>  = \int_0^\infty A(s) e^{-sx}ds\,.
\ee
Note that we are specialising in the modes to the right of the obstacle, but an analogous formula obviously exists for the modes on the left.
We are interested in the function $A(s)$ which describes the density of SCM in the complex momentum plane. We can calculate this using the usual Bromwich inversion formula
\be
A(s)  = \frac{1}{2\pi i}\int_{\gamma - i\infty}^{\gamma + i\infty} \left<O_\phi (x)\right>e^{sx}dx\,,
\ee
substituting the form of $ \left<O_\phi (x)\right>$ for black Janus, i.e. Eq. \eqref{Janusvev}.
We start with the finite-temperature case, which results in a sum of delta functions,
\be
A(s)  = \sum_{n=0}^\infty A_n^+ \delta\left(s-s_n  \right)\,,\qquad \textrm{with}\qquad s_n =  4\pi T(1+n)
\ee
for $A_n^+$ as defined above. Evidently we have recovered a discrete infinity of modes. Plugging this back into \eqref{eq.OphiLaplace}, we find
\be
\left<O_\phi\right>  = \sum_{n=0}^\infty A^+_n e^{-s_n x} =  \sum_{n=0}^\infty A^+_n e^{ik^+_n x}\,.
\ee
 In other words, we recover the discrete set of purely imaginary SCM ascending the positive imaginary momentum axis of \eqref{eq.JanusSCMRight} by means of the Laplace transform.
Repeating now the same procedure for the zero-temperature expression for $ \left<O_\phi (x)\right>$ we find
\be
A(s) = 2\frac{\sqrt{{\sf m}}}{1+{\sf m}} s\,,
\ee
i.e. a continuum of modes emanating from $s=0$. This spectral density, translated into the complex $k$ plane, corresponds to a continuum of modes lying on a branch cut along the positive imaginary $k-$axis emanating from a branch point $k=0$. Let us finally remark that this coincides, as expected, with the naive continuum limit in Eq. \eqref{JanusSCMs}, i.e. taking $T\rightarrow 0$, and defining a continuum variable $s = T n$. This analytic structure is illustrated in \autoref{fig.polesToBranch}.

\section{Nonlinear solutions: steady states without Killing horizons}\label{sec.NonlinearNumerics}
In a previous publication \cite{Sonner:2017jcf}, two of the authors constructed a class of non-equilibrium steady states, and pointed out the relevance of SCM to their spatial asymptotic behavior. In this section, we give more details about the numerical aspects of the construction.

In what follows we will construct a holographic example of a nonlinear non-equilibrium steady state. Our goal is twofold: on the one hand we wish to demonstrate the role played by the spatial collective modes that have been explored in this paper, and on the other, we wish to supply more detail on the actual numerical construction employed to obtain the fully non-linear solutions. Specifically we will demonstrate that the approach to equilibrium at long distances is governed by the SCM.

We construct non-Killing black brane solutions corresponding to flow-past-obstacle steady states in the dual field theory. Solutions of this type, stationary quenches\footnote{Other interesting examples of non-Killing black holes are provided by the flowing solutions of \cite{Fischetti:2012vt}. Nonlinear solutions with time-dependent horizons have been studied earlier, for example in the work of \cite{Chesler:2008hg}.}, have been constructed before, \cite{Figueras:2012rb}. Indeed our results for this section draw on the methods introduced in \cite{Figueras:2012rb}. Here we shall elaborate on the numerical techniques we used, which extend those of \cite{Figueras:2012rb} to include transverse velocity and supersonic asymptotic flow velocities. This extension is necessary so that we may compare to our transverse and supersonic collective modes.

\subsection{Numerical method\label{sec.numericalmethod}}
One way to understand the method is to consider a steady state formed by waiting for a time-dependent process to settle down. This is an inefficient way to access the steady state, and offers little control over which steady state is reached, since the resulting steady state at late times will have its moduli governed by the chosen initial conditions. Specifically, the moduli are a set of numbers which label the flowing solutions, and so for instance dictate the left and right asymptotic energy densities, velocities and incident angles.  Instead, one may {\it skip to the end} and look directly for the stationary solution. In this case there are no initial conditions to determine which solution is obtained, and so the moduli must be fixed by other means. In this work the moduli are conveniently fixed by imposing additional boundary conditions at points behind the future event horizon, following \cite{Figueras:2012rb}. We find that three boundary conditions are required to fix three moduli, which we think of as an asymptotic flow velocity, an asymptotic energy density and an asymptotic incident angle, and we go into some detail on these choices below. 

We denote our bulk metric as $g_{ab}$ with $a,b= 0,\ldots,3$ and boundary directions labelled by $\mu,\nu = 0,\ldots,2$.
To keep things as simple as possible we use the boundary metric itself, $\gamma_{\mu\nu}$ in order to produce an obstacle,
\be
\gamma_{\mu\nu} = \eta_{\mu\nu} + s_{\mu\nu}(x),\label{boundarymetric}
\ee
which depends non-trivially only on one boundary direction, $x$. As is apparent from the rest of this paper, the interesting spatial behavior we observe is expected to be fully universal \cite{Sonner:2017jcf} and would manifest itself in many other ways of introducing an obstacle. Our present choice of obstacle as a boundary metric deformation allows us to consider a minimal bulk theory of pure gravity in AdS$_4$, with no matter fields required.
To deal with the usual gauge issues we adopt the Einstein-DeTurck or `generalised harmonic' equations \cite{Headrick:2009pv, Wiseman:2011by}, and we recommend these references for further details. Briefly, one introduces a reference metric $\bar{g}_{ab}$ and a vector $\xi^a = g^{bc}\left(\Gamma^a_{bc} - \bar{\Gamma}^a_{bc}\right)$ and modifies the equations of motion,
\be
R_{ab} - \nabla_{(a}\xi_{b)} + 3 g_{ab} = 0.\label{einstein}
\ee
To define our reference metric we first introduce Schwarzschild-AdS$_4$ boosted by a 2-velocity $\boldsymbol{\beta}$ with corresponding 3-vector $u^\mu = \frac{1}{\sqrt{1-\delta_{ij}\beta^i\beta^j}}\left(1, \boldsymbol{\beta}\right)$, written in ingoing Eddington-Finkelstein coordinates,
\bea
ds_{Schw.}^2 &=& \frac{1}{z^2}\left(-f(z) (u_\mu dx^\mu)^2 + 2 u_\mu dx^\mu dz + \Delta_{\mu\nu}dx^\mu dx^\nu\right), \label{metricS}\\
f &\equiv& 1- \frac{z^3}{z_h^3}
\eea
where $\Delta_{\mu\nu} \equiv \eta_{\mu\nu} + u_\mu u_\nu$.
This metric also introduces the holographic coordinate $z$ chosen such that the conformal boundary is defined by the double-pole at $z=0$.
We choose our reference metric to be this solution manually adjusted to match the boundary metric choice \eqref{boundarymetric} in the following way, 
\bea
\bar{ds}^2 = ds_{Schw.}^2 + z^{-2} s_{\mu\nu}(x) dx^\mu dx^\nu,\label{refmetric}
\eea
which is not a metric which solves the Einstein equations in general. We ensure that the source components are orthogonal to $u^\mu$,
\be
s_{\mu\nu} = \Delta_\mu^{~\rho}\Delta_\nu^{~\sigma}\mathcal{S}_{\rho\sigma}.
\ee
In this way $u^\mu$ is still unit-norm, $\gamma_{\mu\nu}u^\mu u^\nu = -1$, despite the presence of the obstacle, $s_{\mu\nu}(x)$.
For concreteness we adopt a particularly simple choice of source, 
\be
s_{\mu\nu} = \mathcal{S}_{\mu\nu} = s(x) n_\mu n_\nu
\ee
where $n_\mu = \left(\beta_x^2+\beta_y^2\right)^{-1/2}(0, -\beta_y, \beta_x)$ with a Gaussian choice
\be
s(x) = A e^{-B x^2}
\ee
and free parameters $A,B$. At this point we also introduce the angle of incidence parameter in the reference metric as, 
\be
\tan\theta = \beta^y/\beta^x.
\ee
We emphasise that $\beta^i, z_h$ and $\theta$ are simply reference metric parameters, and do not directly correspond to a final velocity, energy density or angle of incidence in the resulting solution. 

Finally, we find it convenient to factor out the leading divergence in the bulk metric, defining instead $h_{ab}$ through,
\be
h_{ab}(z,x) \equiv z^2 g_{ab}(z,x).
\ee

\subsubsection{Implementation details}
\begin{figure}[h]
\begin{center}
\includegraphics[width=0.5\textwidth]{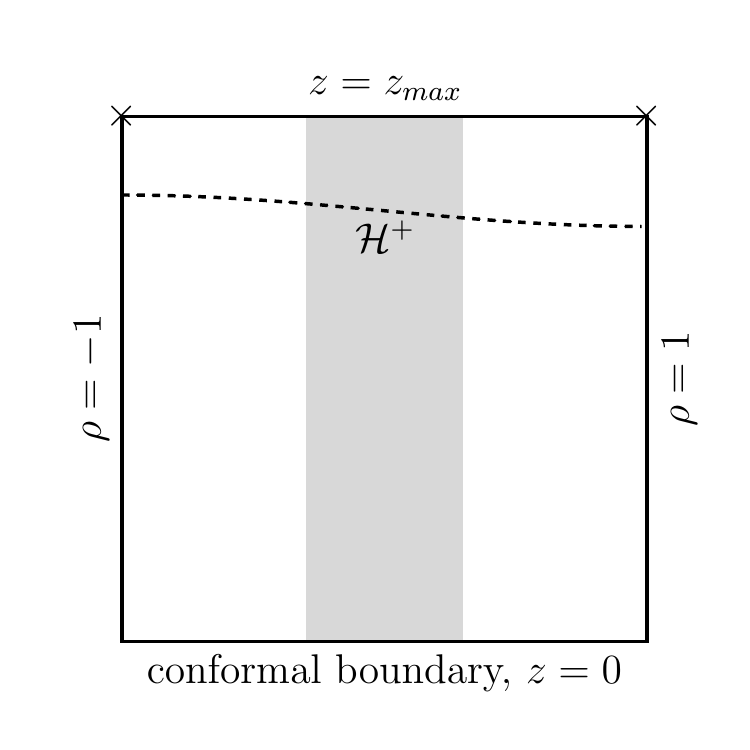}
\caption{A sketch of the domain used for the numerical construction of holographic non-equilibrium steady states. The spatial field theory direction is compactified using coordinate $\rho$ as defined in \eqref{rhodef}. The `$\times$'s mark the points where boundary conditions are imposed to fix moduli of the solution. The gray filled region schematically indicates the presence of the obstacle, in the sense that far to the left or to the right the bulk is described by an equilibrium solution with spatial collective modes. \label{domain}}
\end{center}
\end{figure}

To aid the construction of flows that are infinitely extended and inhomogeneous in the $x$ direction, we compactify using a coordinate $\rho$, 
\be
x = \frac{\rho/\ell}{1-\rho^2} \label{rhodef}
\ee
with $\rho \in [-1,1]$.
The other non-trivial direction is labelled by the holographic coordinate $z \in [0,z_{max}]$ with the conformal boundary of AdS at $z=0$ and the set $z=z_{max}$ should be entirely behind the future event horizon of the solution, ${\cal H}^+$. The parameter $\ell$ allows us to adjust the overall scale of the grid relative to the characteristic size of the inhomogeneities. In practice, typical choices of $\ell$ are ${\cal O}(1)$ in our simulations. Finally there is no dependence in the remaining two directions,  $t,y$. The resulting domain is shown in \autoref{domain}.

We use Dirichlet boundary conditions to fix $h_{ab}$ to be the conformal boundary metric \eqref{boundarymetric} at $z=0$ for all $\rho$.
As discussed we must also introduce data which fix the moduli of the solution, as in \cite{Figueras:2012rb}. This is introduced as additional Dirichlet data behind the horizon in the corners of the grid, at $z=z_{max}, \rho = \pm 1$ for some metric components. Roughly speaking we have three moduli fixing boundary conditions because we have three solution moduli: an energy density, a velocity and an angle of incidence for the flow. Specifically, at these points we set $h_{ab} = z^2 \bar{g}_{ab}$ as follows:
\bgroup
\begin{center}
\begin{tabular}{@{\hspace{1em}}l@{\hspace{1em}}|@{\hspace{1em}}l@{\hspace{1em}}|@{\hspace{1em}}l@{\hspace{1em}}}
asymptotic $v$ &  upstream corner &  downstream corner\\
\hline
subsonic & $h_{tt}, h_{ty}$ & $h_{tz}$\\
supersonic & $h_{tt}, h_{ty}, h_{tz}$ & --
\end{tabular}.
\end{center}
\egroup
Whilst the above choices work well in practise, in the sense that the numerical method converges to a solution to the Einstein equations, we do not have a rigorous understanding of why these particular choices work where certain others do not.
For instance, the choice we arrived at for the supersonic case amounts to a specification of all three moduli in the upstream asymptotic region; it would be interesting to understand the connection to the character of the corresponding problem in fluid dynamics. 
Finally we impose Neumann boundary conditions on all fields along the remaining points of the $\rho = \pm 1$ edges. The remaining points at $z=z_{max}$ are left free to obey the equations of motion.

We utilise a regularly spaced discretisation of $\rho,z$ with $N_\rho, N_z$ grid points respectively, taking $N_\rho = 4 N_z$. We adopt sixth-order finite difference approximations of the derivative operators. The resulting system of equations is solved iteratively using the Newton method. The Jacobian is computed numerically, utilising a second-order centre-difference stencil for taking derivatives of the equations, which is slower to compute but results in considerably better convergence of the Newton method than the first-order finite difference. Computing the Jacobian can be sped up by restricting the difference computation to only the affected areas, i.e. roughly speaking in a stencil-sized box around the varied grid point. The resulting sparse linear system is then solved directly using the LU-factorisation algorithm provided by \verb!UMFPACK!\cite{umfpack}.

The initial guess for the Newton method is taken to be the reference metric, together with a low amplitude ($A$) source. Once obtained we use small $A$ solutions as initial guess metrics for larger $A$ solutions. In all cases we find it is convenient to start at low resolutions (typically $N_\rho = 80, N_z = 20$), and then use sixth-order-interpolated versions of these as solution guesses for higher resolutions. The interpolated guess converges in one or two Newton steps to a solution. In this fashion we have obtained solutions up to $N_\rho = 520,N_z = 130$, limited ultimately here by the memory required for the direct linear solver, but a resolution that is more than sufficient for our requirements. 

\begin{figure}[h]
\begin{center}
\includegraphics[width=0.6\textwidth]{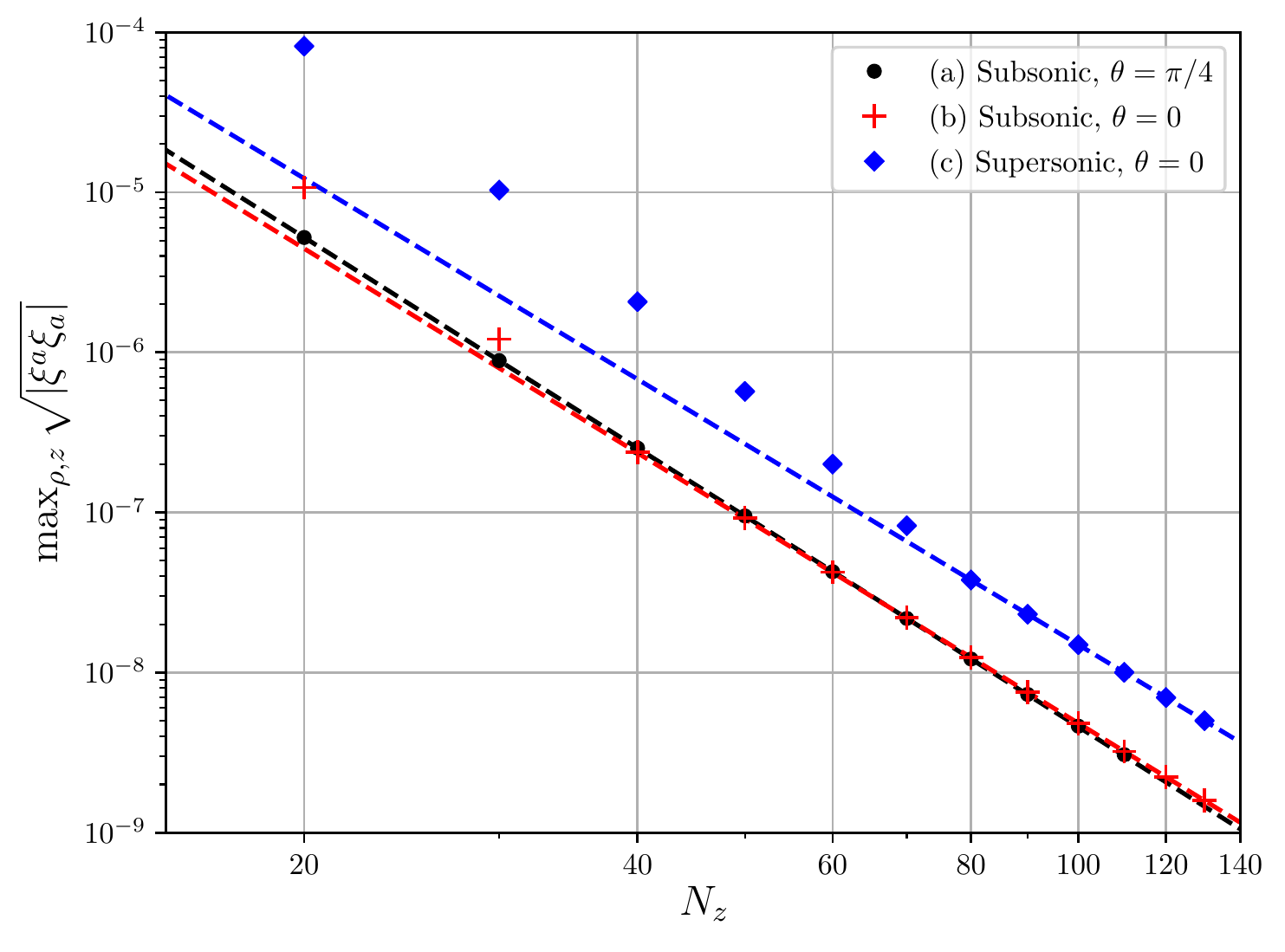}
\caption{Convergence tests of the numerical method for non-Killing black branes describing NESS. We show the approach to the continuum limit of the $\xi^a$ vector, whose maximum value on the numerical grid approaches zero as a power law with order (a) 4.4, (b) 4.2, (c) 4.2.  \label{convergence}}
\end{center}
\end{figure}

For a selection of representative solutions (examples (a), (b) and (c) defined later in section \ref{nonlinear_results}) we have carried out convergence tests, looking at the approach to the continuum limit of the vector $\xi^a$. This is shown in \autoref{convergence} demonstrating results consistent with fourth-order convergence.

\subsection{Results\label{nonlinear_results}}
\begin{figure}[h]
\begin{center}
\includegraphics[width=0.9\textwidth]{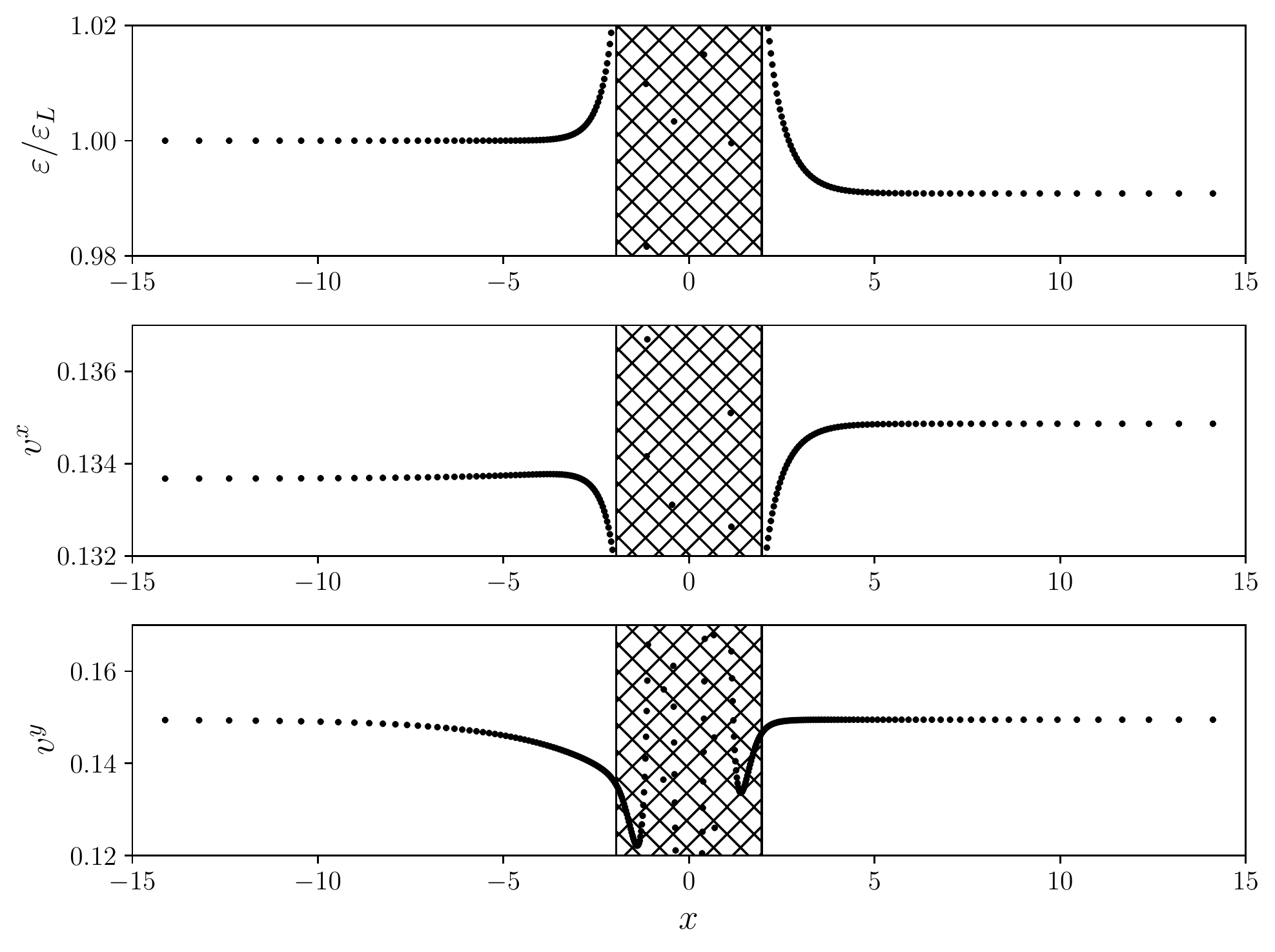}
\caption{Flow profiles for solution (a): subsonic-to-subsonic non-equilibrium steady flow at incidence parameter $\theta = \pi/4$. In this case the fluid is rarefied, refracted and sped up by the obstacle. Note that $v^y_L = v^y_R$. The tails visible in these spatial profiles are the spatial collective modes.\label{flowprof_shear}}
\end{center}
\end{figure}

\begin{figure}[h]
\begin{center}
\includegraphics[width=0.9\textwidth]{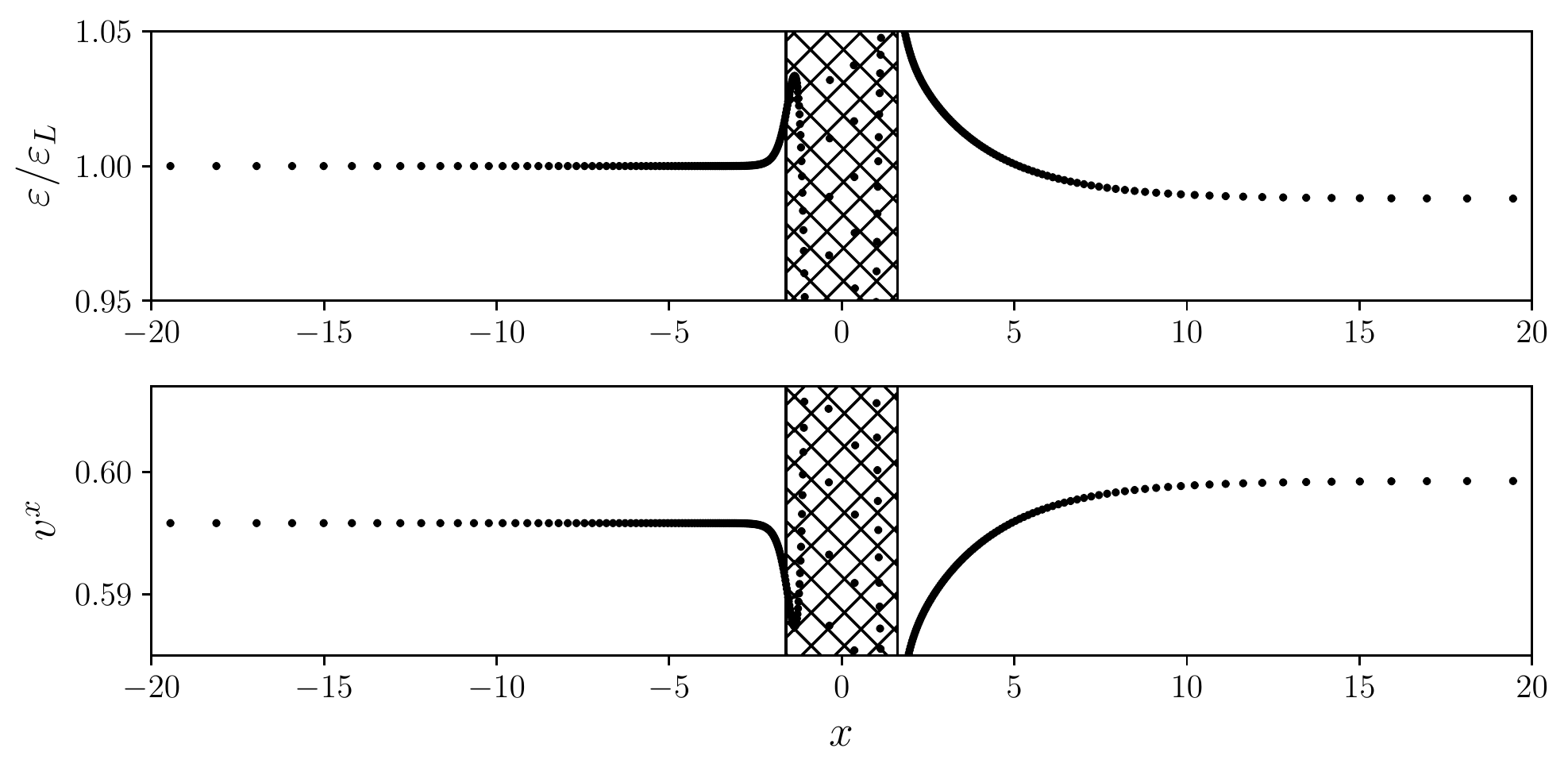}
\caption{Flow profiles for solution (b): subsonic-to-subsonic non-equilibrium steady flow at normal incidence. Here the fluid is rarefied and sped up by the obstacle. The tails visible in these spatial profiles are the spatial collective modes.\label{flowprof_sub}}
\end{center}
\end{figure}

\begin{figure}[h]
\begin{center}
\includegraphics[width=0.9\textwidth]{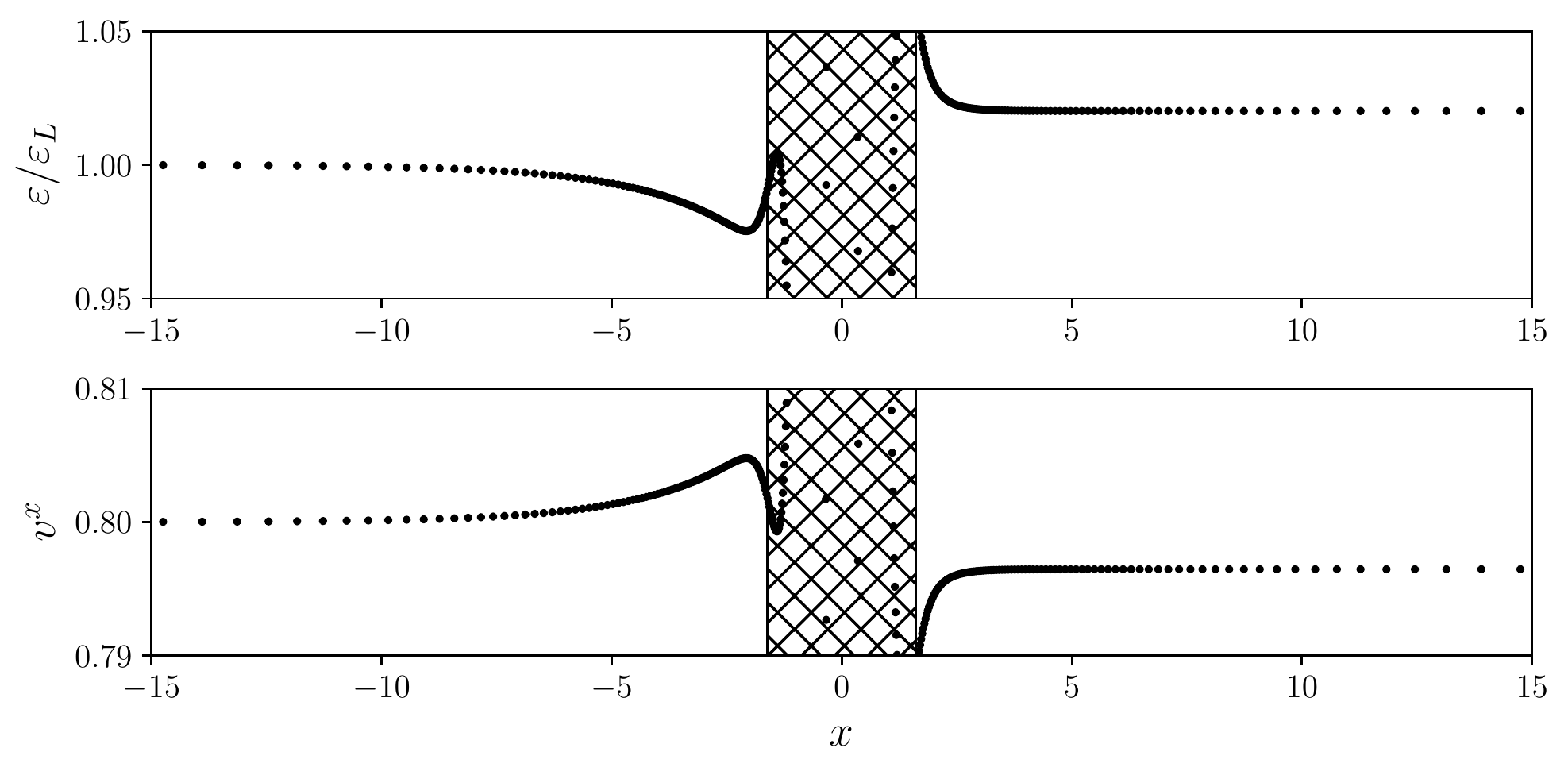}
\caption{Flow profiles for solution (c): supersonic-to-supersonic non-equilibrium steady flow at normal incidence.  Here the fluid is compressed and slowed through its encounter with the obstacle. The tails visible in these spatial profiles are the spatial collective modes. \label{flowprof_sup}}
\end{center}
\end{figure}

With solutions obtained our next task is to interpret the results from the CFT perspective. Our first task is to read off the one-point function of the CFT stress tensor, $\left<T_{\mu\nu}\right>$, from the bulk solutions. The details of this calculation are set out in appendix \ref{app.stresstensor}, and we quote the main result here for convenience,
\be
\left<T_{\mu\nu}\right> = \frac{1}{2}\partial_z^3 h_{\mu\nu}\big|_{z=0} + \frac{\gamma_{\mu\nu}}{z_h^3} + V_{\mu\nu}
\ee
where $V_{\mu\nu}$ is a known term which vanishes outside the obstacle.
With $\left<T_{\mu\nu}\right>$ in hand, we solve the following eigenvalue problem at each point $x$ along the flow, 
\be
\left<T^{\mu}_{~~\nu}\right>U^\nu = -\epsilon U^\mu\qquad \gamma_{\mu\nu}U^\mu U^\nu = -1, \label{eval}
\ee
and define the local flow velocities
\be
v^x \equiv \frac{U^x}{U^t}, \qquad v^y \equiv \frac{U^y}{U^t}.
\ee
Note that it is not always possible to solve \eqref{eval} within the obstacle region where the flow becomes strongly non-linear.
For illustrative purposes we restrict our attention to three different points in moduli space, 
\begin{enumerate}
\item[a)] Subsonic flow at non-zero angle of incidence. The parameters used are, $z_h =  0.975, \beta_x = 0.15, \beta_y = 0.15, A= 1.0, B = 3.0, \ell=0.5$. The corresponding $\varepsilon,v^x,v^y$ are shown in \autoref{flowprof_shear}.
\item[b)] Subsonic flow at zero angle of incidence, $\theta = 0$. The parameters used are, $z_h =  0.975, \beta_x = 0.6, \beta_y = 0.0, A= 0.1, B = 2.9618, \ell=0.5$. The corresponding $\varepsilon,v^x$ are shown in \autoref{flowprof_sub}. 
\item[c)] Supersonic flow at zero angle of incidence, $\theta = 0$. The parameters used are, $z_h =  0.975, \beta_x = 0.8, \beta_y = 0.0, A= 0.1, B = 3.0, \ell=0.5$. The corresponding $\varepsilon,v^x$ are shown in \autoref{flowprof_sup}. 
\end{enumerate}

\begin{figure}[h!]
\begin{center}
\includegraphics[width=0.9\textwidth]{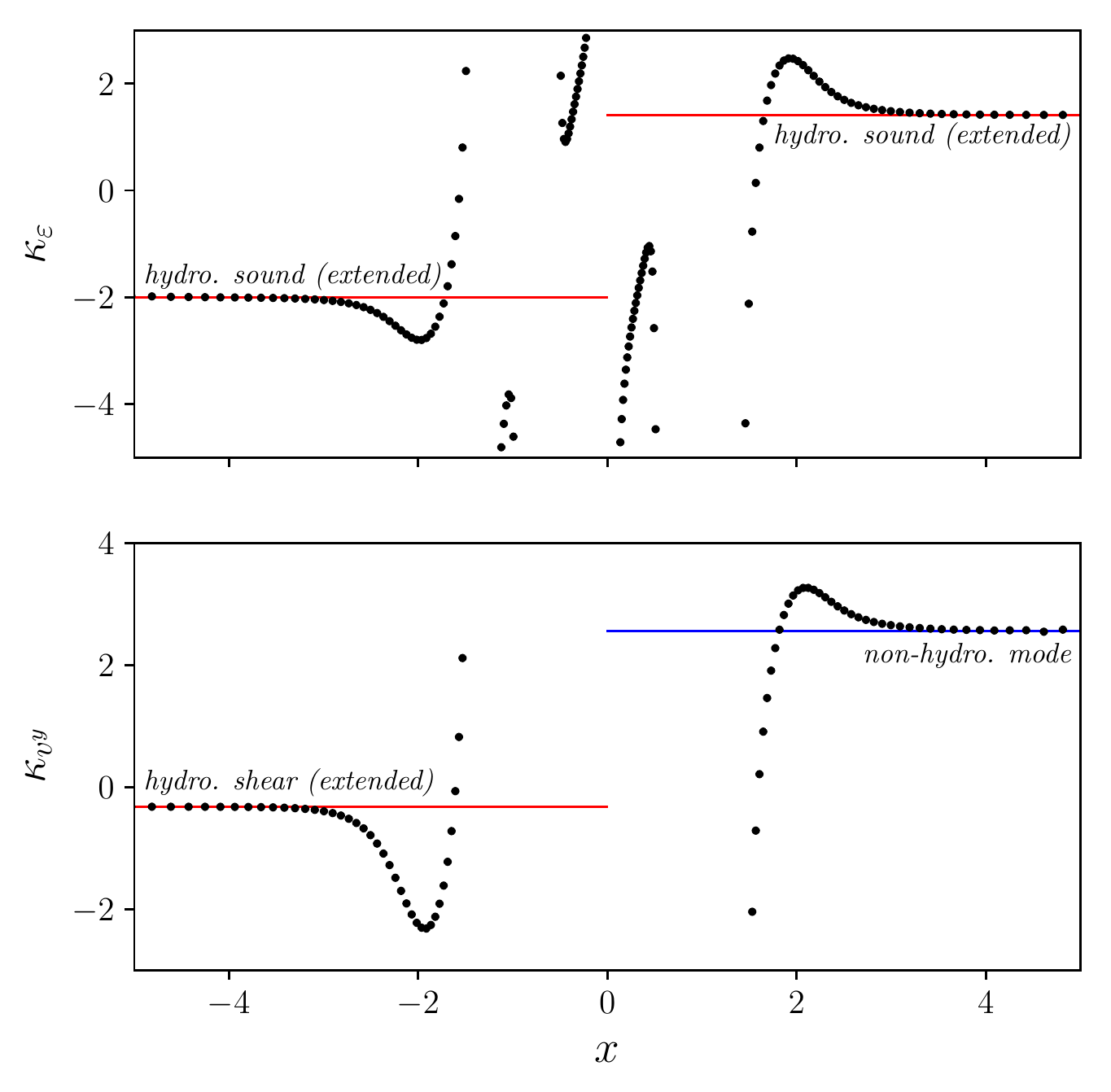}
\caption{Comparing the asymptotics of solution (a) to the spatial collective modes appropriate to each asymptotic equilibrium region. The points are given by derivatives of the spatial profiles of solution (a), according to the definition \eqref{kappadef}. The spatial collective modes are shown by the coloured solid lines. Note that a non-hydrodynamic mode is excited in the transverse channel on the downstream side. \label{matching_a}}
\end{center}
\end{figure}

The imprint of the spatial collective modes is already apparent in the flow profiles of \autoref{flowprof_shear}, \autoref{flowprof_sub}, and \autoref{flowprof_sup}. To demonstrate this more clearly we take derivatives of the data in order to extract the decay length and compare it to the modes of section \ref{sec.RN_SCM}, given the specific asymptotic equilibrium approached. Specifically, for some quantity $f$ we define,
\be
\kappa_f(x) \equiv -\frac{1}{\varepsilon^{1/3}}\frac{\partial_x^2 f}{\partial_x f}. \label{kappadef}
\ee
Then, if the asymptotic functional form of $f$ is given by $f = C + A_k e^{-{\rm Im}k\,x}$, the asymptotic value of $\kappa_f$ will be the wavenumber itself, ${\rm Im}k/\varepsilon^{1/3} = \lim_{x\to\pm\infty} \kappa_f(x)$. A comparison is shown for the solution (a) in \autoref{matching_a}, illustrating the existence of these modes in the flow profile, including a mode which is of non-hydrodynamic origin.

\begin{figure}[h!]
\begin{center}
\includegraphics[width=0.9\textwidth]{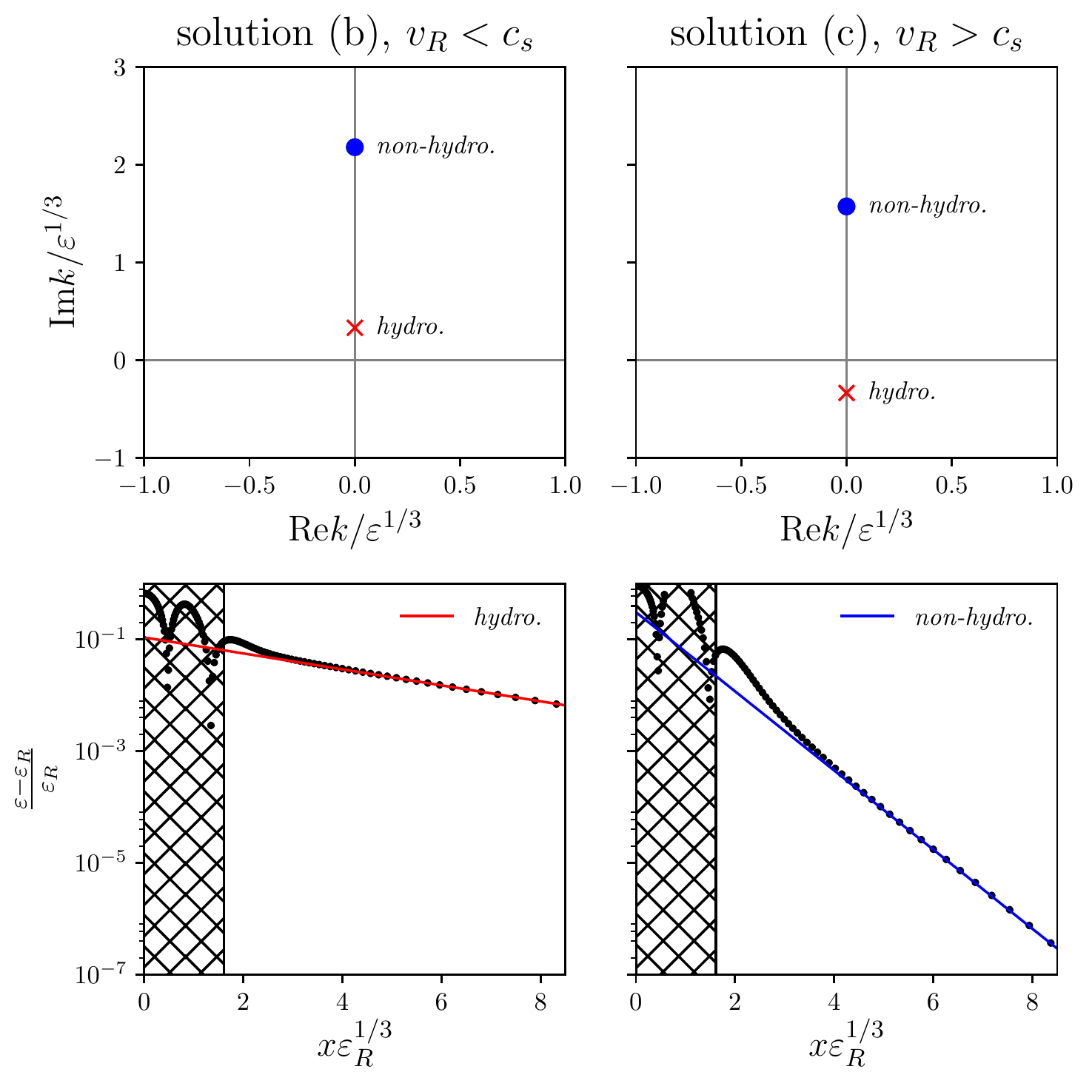}
\caption{Matching the downstream asymptotics of the subsonic flow solution (b) (left column) and the supersonic flow solution (c) (right column) to spatial collective modes. Upper panels show the leading pole structure in the complex-$k$ plane as computed in \cite{Sonner:2017jcf} in the neutral case of relevance here, and in section \ref{sec.RN_SCM} with finite charge density. Lower panels show the asymptotic behaviour of $\varepsilon$ using log-linear axes.  The black dots are the nonlinear solutions and the solid lines are the spatial collective modes. As $v_R$ is increased through $c_s$ the hydrodynamic spatial collective mode (red) transitions from downstream to upstream and the long-range behaviour downstream jumps to the longest non-hydrodynamic mode (blue). \label{matching_bc}}
\end{center}
\end{figure}

Next, we perform a similar analysis in order to investigate the asymptotic behaviour of solutions (b) and (c). In moving from (b) to (c) we move from subsonic to supersonic flows and we expect to see the non-equilibrium phase transition due to a complex-$k$ mode crossing the real axis.
To illustrate this more clearly, we restrict to the downstream side and subtract the asymptotic value. Then on a log-linear plot, the slope will give the appropriate spatial collective mode. This is shown together with the leading pole structure in the complex-$k$ plane for downstream (b) and downstream (c), in \autoref{matching_bc}.

\section{Summary and Discussion}\label{sec.SummaryDiscussion}
Before moving to discuss open issues and future work, let us pause to briefly recapitulate the salient features of this work. In \cite{Sonner:2017jcf} two of us proposed a universal description of a class of nonequilibrium steady states, motivated by holographic duality. This description relies on a set of modes, the spatial collective modes (SCM), defined in the complex momentum plane, which are spacelike cousins of quasinormal modes, used to describe universal equilibration dynamics in holography for systems excited by an explicit time-dependent perturbation. The SCM, instead, describe the spatial relaxation of nonequilibrium steady states, and are excited by spatial inhomogeneities. The behavior away from the source is universal in the sense that it depends only on the theory and the asymptotic configuration that is approached but not on the details of the inhomogeneities.

In section \ref{sec.DefinitionSCM} we gave a description of these modes from the point of view of the boundary field theory, where they appear notably as poles of correlation functions in the complex momentum plane. We then embarked on a range of calculations in order to illustrate SCM, in various interesting contexts. The most complete understanding, unsurprisingly, is obtained in holography. We explicitly construct non-linear solutions (based on the methods described in \cite{Figueras:2012rb}) by forcing the strongly coupled fluid dual to Einstein gravity over an obstacle, modelled by sourcing the boundary metric, and match the asymptotic spatial behavior of such steady state solutions to the leading SCM appearing in the complex momentum plane of the relevant two-point functions. As we have pointed out these are defined by boundary conditions which are regular at the horizon and in one of the nontrivial spatial directions. We have considered the SCM in a variety of systems and have given a general numerical recipe for their calculation. There  are a number of contexts where we can determine the spectrum of modes analytically, notably three-dimensional holography (the BTZ black hole) and the large-D limit of \cite{Emparan:2015rva}. An interesting feature of all cases we considered is that the SCM were always purely imaginary for a neutral fluid, but could acquire non-zero real parts, that is represent damped oscillatory behaviour, at finite chemical potential. We pointed out that there are interesting critical phenomena at the points where pure decay transitions into damped oscillations, producing nonequilibrium phase transitions of the same flavor as those first described in \cite{Bhaseen:2012gg}. Due to the ubiquity of SCM as well as QNM we expect to find examples of such transitions recurring in many more contexts, and indeed we remark that similar phenomena have already appeared in \cite{Withers:2016lft,Brattan:2010pq}. Each of these transitions can be reduced to either a pole collision or an exchange of dominance between poles. In the former case, the vanishing of the real part of the mode proceeds with a critical exponent $1/2$. It would be of interest to develop a general mean-field treatment of these phenomena, in the spirit of the theory of dynamic critical phenomena of \cite{Hohenberg:1977ym}.

A particularly well controlled case arises for three dimensions, where we already pointed out that we can construct the full spectrum of SCM analytically. As it turns out, in this case we were able to find a full non-linear example analytically, namely the black Janus solution of \cite{Bak:2011ga} and we point out that its spatial relaxation on either side of the obstacle is precisely given by the leading SCM. In fact one can go further and show that the entire black Janus solution can be re-expressed as a sum over the full spectrum of SCM of the BTZ black hole, which start out as a linear superposition in the boundary, and construct the in-filling bulk solution as a non-linear backreacted version of the tower of modes. This is reviewed in detail in section \ref{sec.BlackJanus} where we also point out that one can reconstruct the spectrum SCM via an inverse Laplace transform of the boundary expectation values of fields in the black Janus solution of \cite{Bak:2011ga}.

A subset of the SCM appearing this work can be constructed using a hydrodynamic effective theory (which we develop in detail), with decay lengths depending on hydrodynamic transport coefficients, such as $\eta/s$ and various diffusion constants. Interestingly we also exhibited cases where the hydrodynamic effective theory does not capture the leading behavior in one or the other asymptotic direction, and we see direct manifestations of higher SCM, as well as interesting phase transitions between hydrodynamic and non-hydrodynamic fall-offs. The steady state thus encodes in a time independent way the hydrodynamic transport coefficients, which may be read off from the spatial decay properties of certain modes excited by the obstacle. In particular the shear SCM decays with length proportional to $\eta/s$ and is thus a direct probe of the shear viscosity to entropy density ratio of whatever strongly-coupled fluid is set up in such a steady state. Experimental evidence for strongly coupled electron flow has been seen in ${\rm PdCoO_2}$ \cite{moll2016evidence} and graphene \cite{crossno2016observation}. In \cite{Sonner:2017jcf} we have estimated these decay lengths for graphene at charge neutrality as well as ${\cal N}=4$ SYM, and we give a few more details about these analyses here\footnote{We thank an anonymous referee of \cite{Sonner:2017jcf} for encouraging us to produce such an estimate.}. For the transverse mode in first order neutral hydrodynamics the dispersion relation is given by \eqref{neutraltransverse}, leading to a decay length for a flow $v$ at normal incidence to the obstacle ($\theta = 0$), 
\be
\left|{\rm Im}k\right|^{-1} = \frac{\eta}{s} \frac{c^2}{vT}, \label{experimentlength}
\ee
where we have re-introduced the speed of light, $c$. To estimate this length for graphene we utilise a number of existing results in the literature. Firstly, we introduce the Fermi velocity via $c = v_F$ which we take to be $v_F = 10^6 m/s$ as in the experimental results of \cite{Novoselov:2005kj}. Next we utilise the value for $\eta/s$ computed in kinetic theory in \cite{2009arXiv0903.4178M}, i.e. we take $\eta/s \simeq 0.00815 \left(\log T_\Lambda/T\right)^2 \frac{\hbar}{k_B}$ with UV cutoff $T_\Lambda = 8.34\times 10^4 K$.
Finally we must provide $v$ and $T$ for the experiment of interest. For flow velocities at around $v\simeq 10^4 m/s$ as in the setup of \cite{PhysRevB.92.165433} we obtain $\left|{\rm Im}k\right|^{-1} \simeq 0.7 \mu m$ at standard temperature, whilst the experiments of \cite{doi:10.1063/1.3483130} obtain much higher velocities $v \simeq 3\times 10^5 m/s$ giving rise to $\left|{\rm Im}k\right|^{-1} = 15 nm$ at $T\simeq 400K$. For comparison we may consider the strongly interacting ${\cal N}=4$ SYM plasma with $c=3\times 10^8 m/s$, under the same conditions, finding $\left|{\rm Im}k\right|^{-1} \simeq 2 cm$ and $\left|{\rm Im}k\right|^{-1} \simeq 0.4mm$ in each case. The main difference between the decay lengths in the two theories arises from the quadratic scaling with the speed of light in \eqref{experimentlength}, rather than the minor differences in $\eta/s$.

Let us now move on to a discussion of interesting open issues as well as promising directions for future work. Throughout this work we have emphasized parallels between the SCM here and their timelike cousins, the QNM. Let us now address some important differences. An important conceptual point concerns the issue of causality. From our construction one may get the impression that there really exists a notion of a `spatially retarded' correlation function as defined in \ref{sec.DefinitionSCM}, in complete analogy to the temporally retarded, i.e. causal, one. This is not so, and our construction should be seen more as a convenience in order to exhibit a particular set of modes in the field-theory, essentially by defining correlation functions that are analytic in either the upper or lower half complex $k$ plane. Note that in the QNM context, for a stable system, the analyticity in the complex $\omega$ plane really is dictated by causality; a system reacts to a disturbance after the quench is applied, and does not have a way of `knowing' that it will be perturbed before it actually gets hit. In other words, only modes decaying towards the future are physically relevant and therefore only retarded correlation functions enter the discussion. By contrast, for the `spatial quench' considered here, the system does exhibit both modes that decay toward positive $x$ as well as negative $x$ (although these modes are generically different from one another). This becomes clear if we consider such a steady state as something that is formed at late times via a time dependent process, since in this case all parts of the system have had causal contact with one another regardless of the flow velocity. Thus, even though for much of the discussion we can think of the SCM as being spacelike analogs of QNM with the special direction $x$ being treated like time, fundamentally the equations both of the dual gravity and the field theory are hyperbolic with respect to the time coordinate $t$ and thus the above causal restrictions apply. 

The distinction between QNM and SCM is particularly salient for non-relativistic theories, where we cannot define one as an analytic continuation of a boosted version of the other. Note that this was the way we constructed the tower of SCM in three dimensions. In looking for SCM about an asymptotic state with flow velocity $v$ we instead boosted into a frame where the fluid was at rest and solved the QNM dispersion relation for the boosted values of the frequency and momentum \eqref{eq.boostedOmegaK}. The resulting mode necessarily had complex momentum $k$, and could also have been constructed directly by solving the perturbation equations in the lab frame where the fluid has finite velocity $v$. In other words, in the relativistic context we can consider the dispersion relations of linear modes as being defined on a $\mathbb{C}^2$ spanned by both complex $\omega$ and complex $k$ and SCM and QNM are merely two different slices (real $k$ and real $\omega$ respectively) of the more general situation. This will not be true in the non-relativistic context, which illustrates that generically the two really describe two different classes of physical phenomena. For this reason it would be enlightening to explore our construction of SCM in non-relativistic theories, perhaps starting from a hydrodynamic effective treatment and then moving on to a model with non-relativistic holography.

Other future directions of interest relate to decreasing the amount of symmetry in the steady state, either by increasing the co-dimension of the obstacle, or by adding spatial inhomogeneities along the direction(s) of the obstacle. In the most general setups one will likely have to confront the issue of heating and whether a parametrically large steady state region (both in space and time) can be established \cite{green2005nonlinear,green2006current}.

Finally it will be important to study issue of time dependance, both in the sense of establishing the steady state from an initial equilibrium state, say by gradually switching on the obstacle, and in the sense of stability to perturbations of the steady state itself (this latter question has already been considered for steady states in ideal hydrodynamics by \cite{Fischetti:2016tek}).

\section*{Acknowledgements}
It is a pleasure to acknowledge helpful discussions with A. del Campo, N. Cooper, J. Gauntlett, A. Green, G. Policastro, K. Schalm, U. Schollw\"ock and T. Wiseman over the course of this work. This work has been supported by the Fonds National Suisse de la Recherche Scientifique (Schweizerischer Nationalfonds zur Förderung der wissenschaftlichen Forschung) through Project Grant 200021 162796 as well as the NCCR 51NF40-141869 ``The Mathematics of Physics" (SwissMAP).
\begin{appendix}
\section{Equations of State}\label{app.EqnOfState}
\subsection{Conformal equation of state}
For a conformal system at temperature $T$ and chemical potential $\mu$ in $d$-dimensions we have the following general expression for the equation of state, 
\be
p(T,\mu) = T^d \Phi\left(\frac{\mu}{T}\right)
\ee
From which follows charge, entropy and energy densities using standard thermodynamic relations,
\bea
n(T,\mu) &\equiv& \left(\frac{\partial p}{\partial\mu}\right)_T = T^{d-1}\Phi'\left(\frac{\mu}{T}\right),\\
s(T,\mu) &\equiv& \left(\frac{\partial p}{\partial T}\right)_\mu = T^{d-2}\left(dT \Phi\left(\frac{\mu}{T}\right)-\mu \Phi'\left(\frac{\mu}{T}\right)\right),\\
\epsilon(T,\mu) &\equiv& -p + Ts+\mu n = (d-1)T^d \Phi\left(\frac{\mu}{T}\right).
\eea
After some manipulations, we may obtain the quantities defined in the hydrodynamic analysis in section \ref{sec.HydroSCM},
\bea
\beta_1 &=& \frac{1}{d-1},\\
\beta_2 &=& 0,\\
\alpha_1 &=& \frac{T^{1-d} \Phi'}{(d-1)\left(\Phi'\right)^2 - d \Phi \Phi''}.\\
\alpha_2 &=& \frac{-dT^{2-d} \Phi}{(d-1)\left(\Phi'\right)^2 - d \Phi \Phi''}.
\eea

\subsection{Reissner-Nordstr\"om AdS$_{d+1}$ equation of state}
For a strongly coupled fluid holographically dual to a Reissner-Nordstr\"om AdS$_{d+1}$ black brane, as a solution to the equations of motion of \eqref{EMaction} with $2\kappa^2 = \tilde{g}^2 = L = 1$, has the following conformal equation of state,
\be
\Phi(X) = R(X)^d\left(1+\frac{(d-2)X^2}{2(d-1)R(X)^2}\right)
\ee
where $R$ is the positive solution of
\be
R^2-\frac{4\pi}{d}R - \frac{(d-2)^2 X^2}{2d(d-1)} = 0.
\ee
Using the above definitions this leads to the expressions,
\be
n = (d-2)\mu r_0^{d-2},\qquad s = 4\pi r_0^{d-1}, \qquad \epsilon = (d-1)r_0^d\left(1+\frac{d-2}{2(d-1)}\frac{\mu^2}{r_0^2}\right), 
\ee
where $r_0 \equiv T R\left(\frac{\mu}{T}\right)$ gives the coordinate position of the event horizon in a Schwarzschild coordinate system. For completeness we finish this section by providing the associated first order hydrodynamic transport coefficients for this state,
\be
\eta = \frac{s}{4\pi}, \qquad \zeta = 0, \qquad \sigma = \left(\frac{sT}{\epsilon + p}\right)^2.
\ee
where $\sigma$ was computed in \cite{Hartnoll:2007ip}.

\section{Extracting the holographic stress tensor\label{app.stresstensor}}
Once we have numerically constructed the NESS, as described in section \ref{sec.numericalmethod}, we wish to extract the one-point function of the CFT stress tensor. Holographic renormalisation is readily performed in Fefferman-Graham (FG) coordinates, where the one-point function is given by a term in the near-boundary expansion there.  However our numerical solutions are not obtained in FG coordinates, rather, they are obtained in coordinates defined by $\xi = 0$. To compute the stress tensor using existing holographic renormalisation results \cite{deHaro:2000vlm} we must find the coordinate map which relates the two. 

\subsection{Near-boundary solution in FG coordinates}
We seek the near-boundary solution in Fefferman-Graham form so that we may use existing results for holographic renormalisation. Taking $z,x^\mu$ to be such coordinates, then by definition $h_{z\mu} = 0$ and $h_{zz} = 1$. Additionally, we compute the following near-boundary expansion of the solutions to \eqref{einstein} at $\xi = 0$ as
\be
h_{\mu\nu}(z,x) = \eta_{\mu\nu} + s_{\mu\nu}(x) + h^{(2)}_{\mu\nu}(x) z^2 + h^{(3)}_{\mu\nu}(x) z^3 + O(z)^4
\ee
where the $h^{(2)}_{\mu\nu}$ are given by 
\bea
h^{(2)}_{00}(x) &=& \frac{\beta_x^2}{\beta_x^2+\beta_y^2} \frac{2(1+s)s'' - (s')^2}{8(1+s)^2}\\
h^{(2)}_{0i}(x) &=& 0\\
h^{(2)}_{ij}(x) &=& \frac{\beta_x^2}{\beta_x^2+\beta_y^2} \frac{2(1+s)s'' - (s')^2}{8(1+s)^2} (\eta_{ij} + s_{ij}(x))
\eea
Aside from constraints imposed by the conformal and diffeomorphism Ward identities, the $h^{(3)}_{\mu\nu}(x)$ are unconstrained, and in holographic renormalisation yield the one-point function of the stress tensor, \cite{deHaro:2000vlm},
\be
\left<T_{\mu\nu}\right> = 3 h^{(3)}_{\mu\nu}.
\ee

\subsubsection{Converting from FG to $\xi = 0$ coordinate system}
From the above analysis we have an expression for the CFT stress tensor in terms of data in a FG near-boundary expansion. Let us map from FG to a new set of coordinates near the boundary
\bea
t &\to & t + \sum_{n=1} T_n(x) z^n \\
x &\to & x + \sum_{n=1} X_n(x) z^n \\
y &\to & y + \sum_{n=1} Y_n(x) z^n \\ 
z &\to & z + \sum_{n=1} Z_n(x) z^{1+n}  
\eea
We take the line element in near-boundary FG expansion and apply these coordinate transformations. The resulting near-boundary metric is then used to compute the vector $\xi$. For the first few couple of orders we find the following choices render $\xi =0$,
\bea
T_1 = \gamma, \quad X_1 = \gamma \beta_x, \quad Y_1 = \gamma \beta_y,\quad Z_1 = \frac{\gamma \beta_x s'}{16(1+s)}\\
T_2 = 0, \quad X_2 = 0, \quad Y_2 = 0, \quad Z_2 = \frac{\gamma^2 \beta_x^2 (64(1+s)s'' - 57 (s')^2}{1024 (1+s)^2}.
\eea
We continue in this way to reach the order at which the stress tensor enters, with expressions that are too cumbersome to present here.  The resulting metric in $\xi = 0$  coordinates contains the sought after data defined through the FG expansion, i.e. $\left<T_{\mu\nu}\right>$, and so by identifying where these terms appear allows us to extract the stress tensor from a bulk solution in $\xi = 0$ coordinates. We find, 
\be
\left<T_{\mu\nu}\right> = \frac{1}{2}\partial_z^3 h_{\mu\nu}\big|_{z=0} + \frac{\gamma_{\mu\nu}}{z_h^3} + V_{\mu\nu}
\ee
where $V_{\mu\nu}$ vanishes when $s'=s''=s'''=0$. Note that this expression depends explicitly on the quantities $z_h, \beta_i$ which are introduced by the gauge $\xi = 0$ through the reference metric.

\end{appendix}

\bibliographystyle{utphys}
\bibliography{NESS}{}

\end{spacing}
\end{document}